%% file: main.tex
\documentclass[letterpaper,twocolumn,10pt]{article}
\usepackage{usenix-2020-09}

\usepackage{preamble}
\usepackage{aliases}
\usepackage{amsmath}

\begin{document}

\title{Nanosecond Precision Time Synchronization for Optical Data Center Networks}

\author{
\rm{Yiming Lei$^{\text{1}}$ \enskip
    Jialong Li$^{\text{1}}$ \enskip
    Zhengqing Liu$^{\text{2}}$ \enskip
    Raj Joshi$^{\text{3}}$ \enskip
    Yiting Xia$^{\text{1}}$ \enskip}\\
    {$^{\text{1}}$Max Planck Institute for Informatics\enskip$^{\text{2}}$Imperial College London\enskip$^{\text{3}}$Harvard University}
}

\maketitle

\input{abstract}

\input{intro}
\input{background}

\input{design_overview}

\input{design_prep}

\input{design_operation}

\input{testbed}

\input{eval}

\input{conclusion}

\newpage
\bibliographystyle{plain}
\bibliography{ref}

\clearpage
\input{appendix}

\end{document}

%% file: abstract.tex
\begin{abstract}

Optical data center networks (\dcns) are renovating the infrastructure design for the cloud in the post Moore's law era. The fact that optical \dcns rely on optical circuits of microsecond-scale durations makes nanosecond-precision time synchronization essential for the correct functioning of routing on the network fabric. However, current studies on optical \dcns 
neglect the fundamental need for accurate time synchronization. In this paper, we bridge the gap by developing Nanosecond Optical Synchronization (\name), the first nanosecond-precision synchronization solution for optical \dcns general to various optical hardware. \name builds clock propagation trees on top of the dynamically reconfigured circuits in optical \dcns,
allowing switches to seek better sync parents throughout time. It predicts drifts in the tree-building process, which enables minimization of sync errors.
We also tailor today's sync protocols to the needs of optical \dcns, including reducing the number of sync messages to fit into short circuit durations and correcting timestamp errors for higher sync accuracy. Our implementation on programmable switches shows 28ns sync accuracy in a 192-ToR setting.

\end{abstract}

%% file: intro.tex
\section{Introduction}\label{sec:intro}

Optical data center networks (\dcns)
are arising as the solution to high-speed, scalable, economical, and power-efficient cloud network infrastructure
~\cite{OSA, WaveCube, MegaSwitch, Helios, Quartz, Mordia, cThrough, ProjecToR, Flat-tree, ShareBackup, OmniSwitch, RotorNet, Sirius, Opera, JupiterEvolving}.
However, optical \dcns work in a fundamentally different way than traditional \dcns, making time synchronization not only a requirement of cloud applications but also a prerequisite for the correct functioning of routing.

\begin{figure}[tbp]
    \centering
    \includegraphics[width=\linewidth]{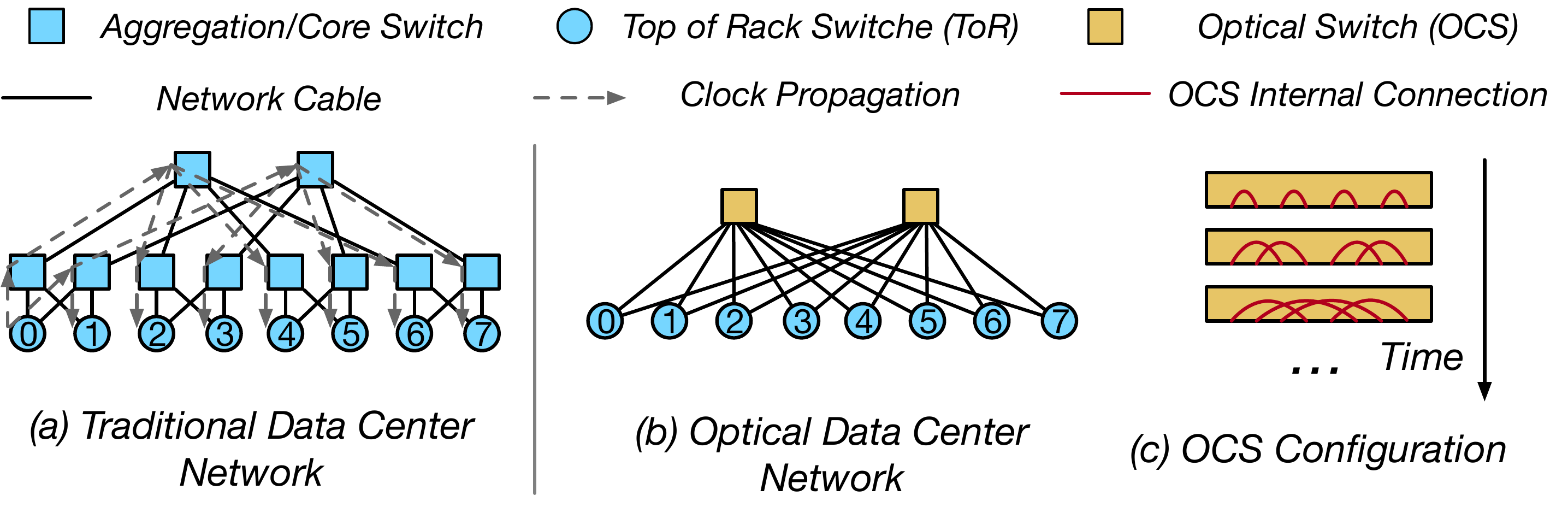}
    \figcap{Traditional vs. optical \dcns.}\label{fig:dcn_arch}
    \vspace{-0.05in}
\end{figure}

Optical \dcns are \textit{circuit-switched}, where optical circuit switches (\ocses)
establish optical circuits
between Top-of-Rack switches (ToRs) before data can be transmitted (Fig.~\ref{fig:dcn_arch}b). An optical controller reconfigures the \ocses throughout time to interconnect different ToR pairs for network-wide connectivity. As a result, ToRs in an optical \dcn need to be synchronized with the optical controller, so that they can transmit data in accordance with the circuit availability.

\begin{figure}[tbp]
    \centering
    \includegraphics[width=\linewidth]{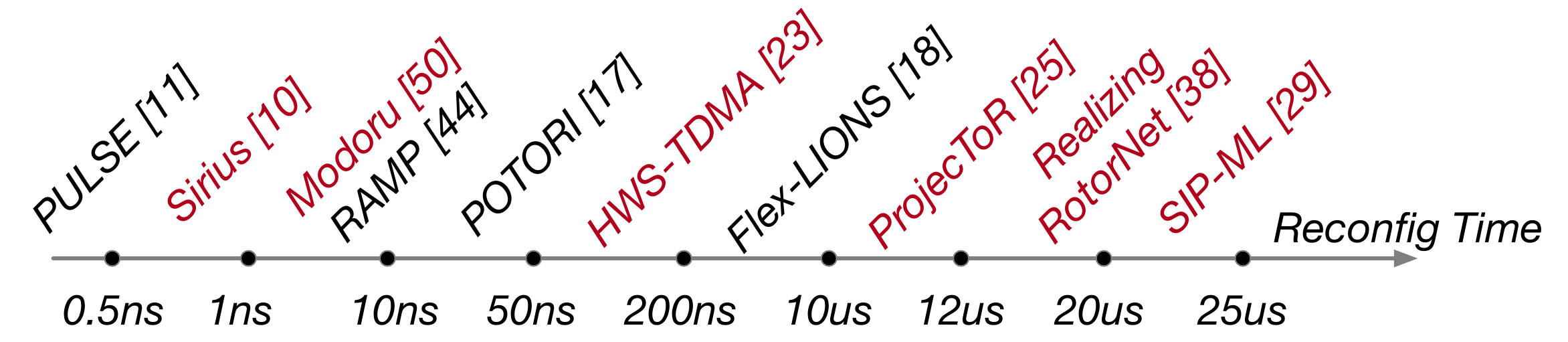}
    \figcap{Optical technologies with $\mu s$- and $ns$-scale reconfiguration delays; \textcolor[RGB]{177, 0, 28}{red} ones demonstrated to be manufacturable.}\label{fig:ocs_hardware}
\end{figure}

The ultimate goal of optical DCNs is to achieve packet-granularity reconfiguration like electrical packet switches, driven by bursty and short flows dominating cloud applications. Recent advances in optical hardware and network fabric, as shown in Fig.~\ref{fig:ocs_hardware}, bring us closer to this vision with microsecond-, nanosecond-, and even sub-nanosecond-scale reconfiguration delays, and many such technologies have been demonstrated to be manufacturable~\cite{PULSE,Sirius,Modoru, RAMP,HWS-TDMA-1,HWS-TDMA-2,POTORI,Flex-LIONS,ProjecToR,RotorNet,SiP-ML}. There has also been increasing vendor support for optical \dcns from the industry. For example, Intel Tofino2 has released the ``decouple mode’’ enabling switch ports to be continuously up during circuit reconfiguration to avoid port flapping from frequent circuit ups and downs~\cite{tofino2}. At this time granularity, the circuit duration can be as short as microseconds or sub-microseconds, necessitating \textit{accurate synchronization at the nanosecond level}.

Despite emerging system integration efforts to embrace the new horizon of optical \dcns~\cite{reTCP, TDTCP, Opera, Sirius, Shale, RealizeRotorNet, UCMP}, synchronization remains an under-studied topic. Most networked systems designed for optical \dcns take synchronization as a given, assuming out-of-band synchronization over a low-speed management network, with only microsecond-level achievable sync accuracy~\cite{Mordia, OSA, ProjecToR, Corundum, MegaSwitch, SiP-ML}; while others rely on specialized optical hardware to extract the sync signal~\cite{Sirius}, which is hard to generalize to other optical architectures.
The lack of accurate synchronization leads to the transitional solution of coarsely reconfigured optical \dcns, which function as a static network most of the time and reconfigures the circuits at hours to days granularity to fit the long-term \dcn traffic~\cite{JupiterEvolving, Gemini, Flat-tree, OmniSwitch}. In this way, they spare the need for rigid synchronization and can reuse the existing network stack, but the cost is the large number of circuits to form a connected graph for the network at any time, as opposed to time-shared circuits if they switch fast.

In this paper, we push the frontier of optical \dcns with Nanosecond Optical Synchronization (\name), \textit{the first general nanosecond-precision time synchronization solution for optical \dcns}.
\name features \textit{in-band synchronization} over the high-speed optical circuits, to overcome the accuracy and generality issues of prior art. This approach makes no additional assumption beyond the basic optical network fabric, such as a parallel management network or specific optical hardware, thus leaving future optical \dcn designs open. 
In \name, we use features supported by commodity switches, moving one more step towards realizing optical \dcns end-to-end.

\name faces two challenges compared to synchronization in traditional \dcns.
(1) Sync protocols for traditional \dcns assume a fully-connected network for continual exchange of sync messages~\cite{DPTP, Sundial}, albeit optical \dcns connect ToR pairs only in discrete time periods.
The transient nature of the optical circuits fundamentally limits when and where sync messages can be sent. (2) ToRs are unaware of circuits' presence before synchronization, making it hard to send sync messages at the right time when circuits are available.

We design a \textit{two-phase sync protocol} to address these challenges.
In the \textit{preparatory phase} before the network starts to operate, we reconfigure the optical circuits to form a connected network topology, which acts as a static network to allow existing sync protocols, such as PTP~\cite{ptp}, to pass the clock from the optical controller to the ToRs.
This process is to bootstrap synchronization by giving ToRs initial clocks of coarse accuracy, so that they can identify the presence of circuits in the \textit{operational phase} later, after the network becomes functional. By so doing, the ToRs can exchange sync messages reliably over the circuits to fine-tune the clocks and re-sync them as they drift over time.

We propose \textit{drift-aware synchronization} as an enabling technique for nanosecond-precision synchronization in optical \dcns. A recent drift characterization study has shown that each device's drifts exhibit a stable expectation value, along with a high variance~\cite{graham}. Leveraging this insight, we predict the drifts for each ToR based on the expectation value. 
Because the dynamic circuits in optical \dcns provide ToRs with rich options of sync parents over time, we use the predicted drifts as a guideline to select sync parents with minimum expected sync errors, and the sync process mitigates the drift variance not captured by the initial predictions. This method provides a significant improvement over current drift-agnostic sync protocols~\cite{ptp, Sundial}, and over simplistic drift prediction strategies that solely compensate the drift expectation, neglecting the high variance~\cite{graham}.

We have implemented \name and evaluated its performance on a testbed with an \ocs and 4 Tofino2 switches, 
as well as on a 192-ToR emulation setting. 
The lightweight computation logic of \name can also be deployed on other hardware, such as NICs and FPGA boards.
\name achieves 28ns sync accuracy under production drift statistics. Additionally, synchronization in optical \dcns with \name is even 1.7$\times$ to 3.8$\times$ more accurate than synchronization in traditional \dcns with the latest sync protocols.
These protocols work for static networks only and cannot apply to optical \dcns.
Our drift-aware synchronization scheme, while tailored for optical \dcns, also inspires a rethinking of sync protocol designs in traditional \dcns.
We will open-source \name to promote followup work on this new problem.

\textit{[This work does not raise any ethical issues.]}

%% file: background.tex
\section{Background and Related Work}\label{sec:background}

\para{Optical \dcns.}
As illustrated in Fig.~\ref{fig:dcn_arch}b, an optical \dcn contains a number of (2 in the figure) optical circuit switches (OCSes) in the optical network fabric. The OCSes connect the ToRs and provide dedicated optical circuits between ToR pairs for high-speed transmission of aggregated traffic. Once established, an optical circuit is retained for a fixed interval of time, called a \textit{time slice}, during which the connected ToRs have exclusive use of it, namely no contention with other ToRs. An \textit{optical controller}, typically on a host machine or an FPGA board,
controls the OCSes to reconfigure the optical circuits continually, once per time slice. The sequence of ToR-wise connections associated with their time slices constitutes an \textit{optical schedule}.
The time slice duration of the latest optical \dcn designs are on microsecond\ or even sub-microsecond scale, with today's OCS technologies that reconfigure circuits in several microseconds or even nanoseconds~\cite{Mordia, RealizeRotorNet, Opera, Sirius}.
These designs feature a pre-defined optical schedule that repeats every \textit{optical cycle}, where there is at least one circuit between every ToR pair per cycle.
Fig.~\ref{fig:dcn_arch}c shows different circuit configurations inside an \ocs.

\begin{figure}[tbp]
    \centering
    \includegraphics[width=\linewidth]{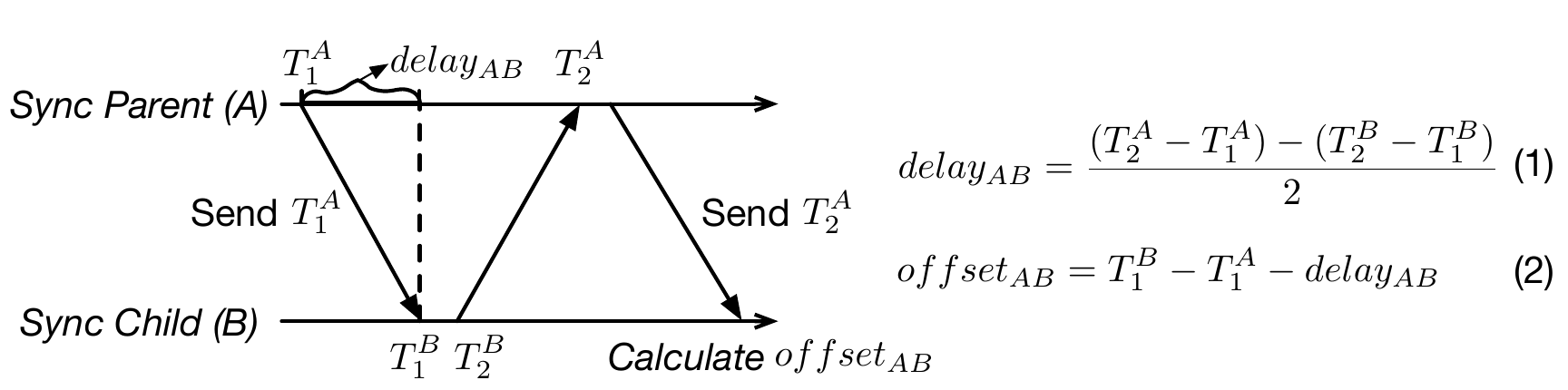}
    \figcap{Message exchanges to synchronize two clocks.}\label{fig:ptp}
\end{figure}

\para{The sync operation.} 
Most sync mechanisms rely on the exchange of sync messages to synchronize the clock of a child node with that of a parent node. Fig.~\ref{fig:ptp} presents this base operation, which involves 3 sync messages~\cite{ptp, DPTP, Sundial}. The first message is for \textit{clock propagation}, where the sync parent A takes the Transmit (TX) timestamp of the message $T_1^{A}$ as its clock and sends it to the child node B.
B notes down the Receive (RX) timestamp of the message $T_1^{B}$ and immediately sends back another message, whose TX timestamp is $T_2^{B}$. Node A captures the RX timestamp of getting the second message $T_2^{A}$ and sends it back to B in the third message. The second and third messages are for \textit{measurement of the propagation delay} between A and B $delay_{AB}$. After collecting the 4 timestamps, node B can calculate $delay_{AB}$ according to Eqn.~1 in Fig.~\ref{fig:ptp}. The clock offset of B, or how much it should adjust its clock relative to A, can be calculated from Eqn.~2. The offset is the difference between B's clock ($T_1^B$) and A's clock compensated by the propagation delay ($T_1^{A}+delay_{AB}$).

\para{Sync in traditional \dcns.}
PTP~\cite{ptp} is the standard sync protocol in traditional \dcns with a static network topology (Fig.~\ref{fig:dcn_arch}a), integrated into switch chips and NICs, achieving nanosecond-level accuracy. Extensions like White-Rabbit\cite{WhiteRabbit} and Sundial~\cite{Sundial} enhance PTP's sync frequency and thus accuracy with specialized hardware, while DPTP~\cite{DPTP} implements it on the data plane of programmable switches. Huygens reduces sync noise on hosts via software~\cite{huygens}, and DTP improves accuracy and lowers sync overhead by delivering sync messages at the physical layer, though it requires replacing network devices~\cite{dtp}.
These solutions assume reliable routing on a static network for sync message exchange, such as using a spanning tree for clock propagation (Fig.~\ref{fig:dcn_arch}a). However, in optical \dcns, synchronization is a prerequisite for routing to align data transmission with circuit availability. Without accurate synchronization in the first place, sync probes will fail when circuits are unavailable.
Graham characterizes clock drifts and corrects the drifts locally~\cite{graham}. As discussed in $\S$\ref{sec:drift-aware}, local compensation alone is insufficient for the accuracy required in optical \dcns and must be supplemented with active synchronization between nodes.

\para{Sync in wireless networks.}
Wireless networks and optical \dcns face distinct synchronization challenges. Wireless networks handle \textit{ad-hoc} come-and-go of nodes at \textit{seconds-to-minutes} intervals, discovering neighbors \textit{online} after nodes join/leave~\cite{Wirelesshart, wirelesssync, wirelesssync2, adhocsync}.
In contrast, optical \dcns follow a \textit{fixed} schedule of known topologies, reconfigured in \textit{microseconds}. This fine time scale requires pre-computation of sync plans \textit{offline}, and the predictable topologies enable global optimization for improve sync accuracy.

\para{Sync in optical \dcns.}
There is a lack of accurate synchronization solutions generalizable to different optical \dcn architectures.
One class of proposals suggest using link up/down events for circuit reconfiguration as the sync signal~\cite{RotorNet, Opera}, but frequent link up/down events from rapid circuit changes would cause port flapping. Other proposals suggest out-of-band synchronization, either through explicit notification to the ToRs in a side channel~\cite{Mordia, ProjecToR, OSA} or via sync protocols on a low-speed management network~\cite{MegaSwitch}. Both approaches have microsecond-scale sync accuracy and thus are mismatched for the ultra-low time slice duration of the latest optical \dcns. The recently released Corundum NIC card has native supports for optical networks, yet their latest report also shows microsecond control precision~\cite{Corundum}.
Sirius has proposed a syntonization approach to extract the clock from data bit streams~\cite{Sirius},
but this solution has not been implemented.
but it is specific to the Sirius optical hardware and cannot generalize to other optical \dcn designs.

\para{Where \name stands.}
\name is an in-band solution for nanosecond precision synchronization
of optical \dcns with arbitrary optical hardware. We focus on construction of clock propagation trees that comply with the dynamic optical circuits, reusing sync message exchange of existing sync protocols, e.g., PTP, as a building block. Specifically, we take DPTP as the underlying sync protocol for \name, because its open-source implementation on programmable switches enables us to modify the tree.
We take PTP and Sundial as our comparison baselines, as tree building is also central to their designs. Sundial, especially, pre-computes backup trees for failure recovery, which has a similar effect as \name's trees evolving with the changing circuits. Graham's finding that drifts are characterizable inspires us to build drift-aware trees through drift profiling. And physical-layer sync of DTP and drift correction of Graham can work as add-ons for \name, which we leave as future work, if the hardware allows.

%% file: design_overview.tex
\section{\name Overview}\label{sec:overview}

In this section, we define the sync problem for optical \dcns and give a high-level overview of the \name design.

\subsection{The Optical Sync Problem}\label{sec:problem}

In an optical \dcn, it is critical to coordinate data transmission at ToRs and hosts with the circuit schedule to avoid packet loss from the network fabric. The optical controller that manages circuit reconfiguration has the \textit{global clock}.
So, this problem boils down to \textit{synchronizing the ToRs with the optical controller},
because hosts are connected to ToRs through static electrical cables, and ToR-to-host synchronization can be trivially done with existing sync protocols, e.g., PTP. Similarly, the optical controller, as a host machine or FPGA board, is connected to a ToR, so it is equivalent to synchronizing with this \textit{master ToR} that is constantly synchronized with the optical controller. 

This sync problem is especially acute for the latest optical \dcns with \textit{microsecond-scale time slice durations} ($\S$\ref{sec:background}), because legacy designs with milliseconds or longer time slices can tolerate the current microsecond precision sync solutions, e.g., explicitly notifying the ToRs of upcoming circuits ($\S$\ref{sec:background}). The sync accuracy should be around two orders of magnitude lower than the time slice duration to ensure correct functionality of optical \dcns~\cite{Sirius}, so the required sync accuracy is in the \textit{low tens of nanoseconds}.

To this end, we aim at \textit{nanosecond-precision ToR-to-ToR synchronization} for optical \dcns. As aforementioned, the major challenge lies in the rapidly changing optical circuits, which fundamentally limit when and where sync messages can be sent. This situation is incompatible with the static spanning tree used for continuous clock propagation in existing sync protocols ($\S$\ref{sec:background}), and more seriously causes a \textit{chicken-egg dilemma}. Although ToRs can be preloaded with the optical schedule and time slice duration, they do not know the starting point of the schedule prior to synchronization. They cannot, hence, send sync messages at the right time, which hinders the acquisition of the clock needed in the first place.

\subsection{Drift-Aware Synchronization}\label{sec:drift-aware}

\begin{figure}[tbp]
\centering
    \includegraphics[width=0.6\linewidth]{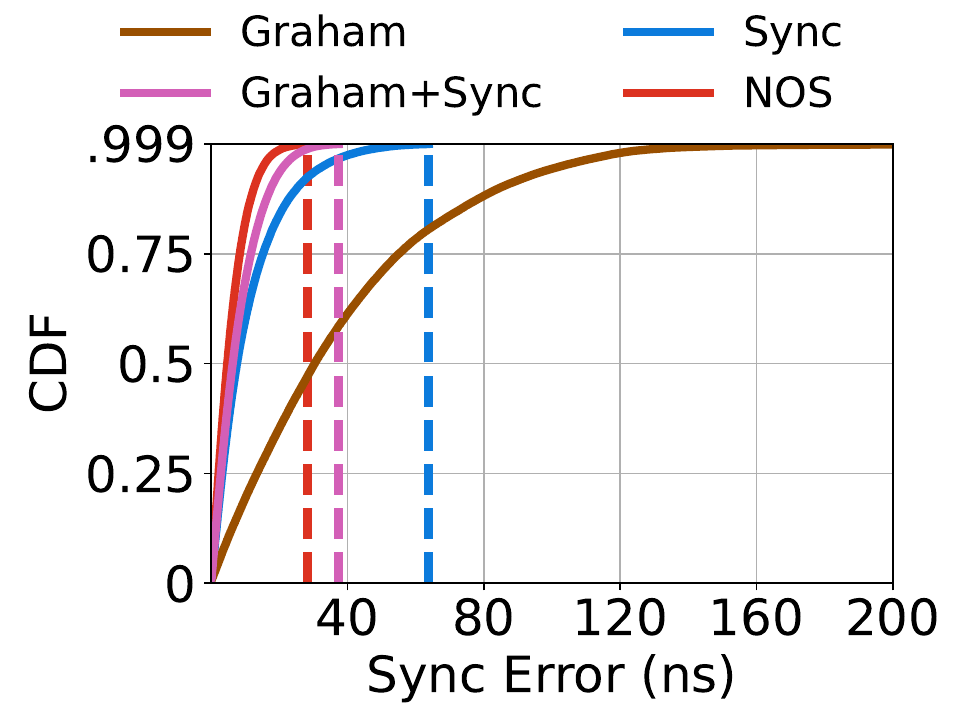}
    \vspace{-0.00in}
    \figcap{Advantage of drift-aware sync in \name.}\label{fig:drift_motivation}
\end{figure}

Classic sync protocols assume drifts are unpredictable~\cite{ptp, ntp}, or at best have a loose bound, e.g., 200ppm\footnote{Parts per million. 1ppm means drifting 1 nanosecond per millisecond.}~\cite{Sundial}. They build drift-agnostic clock propagation trees (Fig.~\ref{fig:dcn_arch}a), where each sync child exchanges sync messages (Fig.~\ref{fig:ptp}) with its sync parent to offset the drift relative to the parent.
Drifts accumulate along the tree, so two conditions must be satisfied to ensure nanosecond-scale sync accuracy.
(1) High sync frequency, e.g., once every 90$\mu s$~\cite{Sundial}, is necessary to update the clock promptly from the master clock. (2) After receiving the clock, a sync parent must forward it immediately to the sync child to minimize the drift accumulation during the clock relay~\cite{Sundial}.
Unfortunately, neither condition applies in optical \dcns due to
the dynamic circuits appearing at discrete times.

A recent drift characterization study, Graham~\cite{graham}, reveals that drifts can be predicted. We include a replication of Graham's key result figure in Fig.~\ref{fig:drift_measurement} of Appx.~\ref{sec:appx_drift} and summarize the findings here.
\textbf{(i)} Different devices present different drift statistics. \textbf{(ii)} The median, or expectation, of drifts per device is stable for each temperature across measurement instances. \textbf{(iii)} Despite the stable expectation, the variance range of drifts under each temperature is large, e.g., exceeding 20ppm.

Based on these observations, Graham predicts each device's drifts by the expectation value, i.e., a constant drift rate, and compensates it locally 
without syncing with the master clock. However, the prediction does not account for the variability of drifts in the large variance range, so it only achieves microsecond-scale sync accuracy.

In \name, we combine the advantages of the sync-based and prediction-based approaches with \textit{drift-aware synchronization}. The dynamic circuits in optical \dcns provide ToRs with rich options of sync parents over time. We predict drifts like in Graham, but instead of local compensation, we use the predictions to find each ToR desirable sync parents with minimum expected sync errors.
Then we perform synchronization, similar to the drift-agnostic protocols, to mitigate the drift variance uncaptured by the prediction. 
In this way, we benefit from a judicious choice of sync parents through drift prediction while avoiding the noise of drift variance.

We validate the effectiveness of our method in Fig.~\ref{fig:drift_motivation}, using the emulation setup in $\S$\ref{sec:eval} of a 192-ToR optical \dcn. 
The drifts per ToR are generated from the median distribution in a production \dcn~\cite{huygens} and variance distribution in Graham~\cite{graham}.
\name is more accurate than local compensation in Graham and drift-agnostic synchronization, which in optical \dcns is syncing each ToR with the master ToR every optical cycle. Notably, the 99.9\textsuperscript{th} percentile tail accuracy of \name is $1.3\times$ better than these two methods combined. Although Graham can compensate drifts between sync operations, it is subjective to errors from the drift variance, while \name incorporates drift awareness in the tree building process to select sync parents with higher expected accuracy.

\subsection{Design Highlights and System Diagram}\label{sec:diagram}

\begin{figure}[tbp]
\centering
    \includegraphics[width=0.8\linewidth]{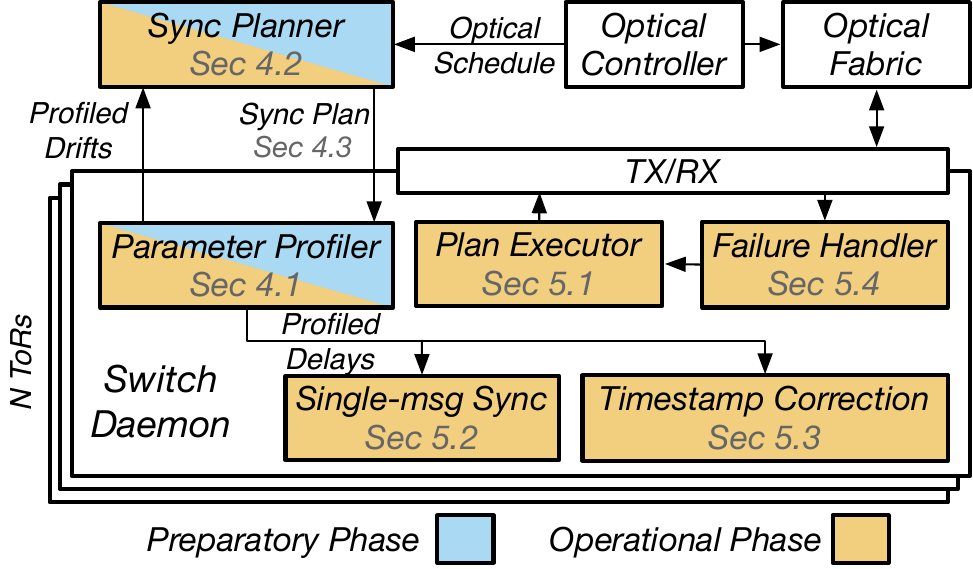}
    \vspace{-0.00in}
    \figcap{\name System Diagram.}\label{fig:system}
\end{figure}

We resolve the chicken-egg dilemma ($\S$\ref{sec:problem}) with \textit{two-phase sync}.
We bootstrap synchronization with an offline \textit{preparatory phase}, where we reconfigure the circuits into a static network topology and run PTP to give the ToRs initial clocks of coarse accuracy. These clocks allow the ToRs to identify time slices in the optical schedule after the network goes online in the \textit{operational phase}, so that they can send sync messages over the expected circuits to fine-tune the clocks.

Fig.~\ref{fig:system} depicts the \name system diagram to realize the idea. Sync planner is the brain of \name. It either co-locates with the optical controller or resides on a host machine that has reliable connectivity with the optical controller through the optical network fabric. There is a switch daemon on every ToR running the functionalities of the preparatory phase ($\S$\ref{sec:preparatory}) and the operational phase ($\S$\ref{sec:operational}).

To enable \textit{drift-aware sync} ($\S$\ref{sec:drift-aware}), the daemons profile the drifts ($\S$\ref{sec:profiling}) in the preparatory phase. The sync planner collects the profiled drifts and incorporates the optical schedule from the optical controller to compute clock propagation trees, known as the sync plan, to minimize the expected sync errors per ToR ($\S$\ref{sec:sync-plan}). 
Then it distributes the sync plan ($\S$\ref{sec:distribution}) to every ToR to execute the plan for actual synchronization ($\S$\ref{sec:enforcement}) in the operational phase.

The daemons also profile propagation delays in the preparatory phase. They are kept local to the ToRs in the operational phase to enable two optimization features:
\textit{single-message sync} to support short time slices ($\S$\ref{sec:single-msg}) and \textit{TX timestamp correction} to improve sync accuracy ($\S$\ref{sec:hop-error-comp}).
The failure handling process ($\S$\ref{sec:failure}) runs in the background and triggers the use of the backup sync plan when failures happen.

The daemons re-profile the drifts in response to significant temperature changes, and 
the sync planner updates the trees and re-distributes the sync plan to the ToRs. This task is performed live during the operational phase, without interrupting the running network.

%% file: design_prep.tex
\section{\name Preparatory Phase}\label{sec:preparatory}

The preparatory phase is to get the prerequisites for the sync process in the operational phase, including (1) the initial clocks of the ToRs to identify the time slices and (2) the clock propagation trees along the optical circuits. We aim to optimize sync accuracy, or minimize sync errors, when building the trees. However, accurate real-time measurement of sync errors is difficult, if not impossible, since it requires the syncing and synced ToRs to share the same clock,
which contradicts the purpose of synchronization in the first place. 
So, we build the trees based on estimated sync errors, which we obtain from profiling key parameters of synchronization, such as drifts and propagation delays. Thus, in the preparatory phase of \name, we profile the sync parameters ($\S$\ref{sec:profiling}), build the trees from the profiled results ($\S$\ref{sec:sync-plan}), and distribute the initial clocks and constructed trees to the ToRs ($\S$\ref{sec:distribution}).

\subsection{Parameter Profiling}\label{sec:profiling}

The sync accuracy is under the mutual influence of \textit{drifts} and \textit{hop errors}.
Drift is the accumulated frequency difference from the master clock over time, while hop error refers to the additional sync error introduced at each hop in the clock propagation tree, caused by sync protocol artifacts and inaccurate ToR timestamping~\cite{DPTP}.
As detailed in $\S$\ref{sec:hop-error-comp}, we observe in Fig.~\ref{fig:sync_system_error} that the inaccuracy of ToR timestamping dominates the hop error. We have developed a timestamp error correction technique that significantly reduces the expectation of hop errors to near zero (median values after correction in the figure), despite
noises from the sync protocol (variance in the figure).

We estimate the sync error by its expectation, which we model as the sum of the \textit{drift expectation} and the \textit{hop error expectation}. Given the near-zero hop error expectation, the sync error can be estimated by the drift expectation.
Recall from $\S$\ref{sec:drift-aware} that Graham reports stable median, or expectation, of drifts per device against temperature, along with non-negligible variance~\cite{graham}. Hence, we can profile the \textit{median value of drifts} per ToR at room temperature, and re-profile it when the temperature varies enough to change the drift expectation.
Then we build clock propagation trees based on the profiled median of drifts (or ``profiled drifts'' for simplicity) and perform synchronization to mitigate the variance of drifts.

We also profile \textit{propagation delays}
as required by single-message sync ($\S$\ref{sec:single-msg}) and timestamp correction ($\S$\ref{sec:hop-error-comp}). Optical \dcns have stable propagation delays, thanks to the fact that ToRs are always directly connected (Fig.~\ref{fig:dcn_arch}b).
\ocses are physical-layer devices without memory buffer, so ToRs connected to \ocses are equivalent to being connected directly through optical fibers (Fig.~\ref{fig:dcn_arch}b). This property excludes queuing and processing delays of intermediate switches. The only remaining source of inaccuracy is measurement noise, which is within a few nanoseconds given hardware timestamps common-place in modern \dcns~\cite{tofino2}.

\para{Drift profiling.} Drifts are intrinsic to physical characteristics of the switch chips thus specific to each ToR. We leverage reconfigurability of optical circuits to connect each ToR to the master ToR in turn (Fig.~\ref{fig:tree-example}a). We maintain each connection sufficiently long for repeated synchronization of each ToR with the master ToR $n$ times every interval $t$ of the time slice duration.
We follow the standard method to calculate the drift between consecutive sync requests~\cite{DPTP}, and we take the median of the collected values as the expectation of the drift per time slice.

\para{Propagation delay profiling.}
Propagation delays need to be profiled per ToR port pair due to different cable lengths.
Once again, we leverage the circuit reconfigurability of optical \dcns to connect every pair of ToR ports, and we measure their propagation delays following Eqn.~1 in Fig.~\ref{fig:ptp}. Similar to drift profiling, we collect $n$ data points and take the median.

\para{Live re-profiling.}
Propagation delays remain stable unless the \dcn is re-cabled. However, changes in temperature, e.g., from cooling malfunctions and switch overheating under heavy traffic~\cite{Sundial}, can cause drifts to shift. We re-profile the median drift when the temperature sensor of a ToR detects a significant temperature change, e.g., over 5$^\circ$C, which, according to Graham~\cite{graham}, is sufficient to alter the drift expectation.
We set the high threshold to avoid frequent re-profiling caused by small temperature fluctuations. Re-profiling is done live without stopping the network. It is the same as profiling, but in the operational phase over the direct circuit to the master ToR, which appears at least once per optical cycle.

\subsection{Sync Plan Generation}\label{sec:sync-plan}

\begin{figure*}[t]
    \centering
    \includegraphics[width=\linewidth]{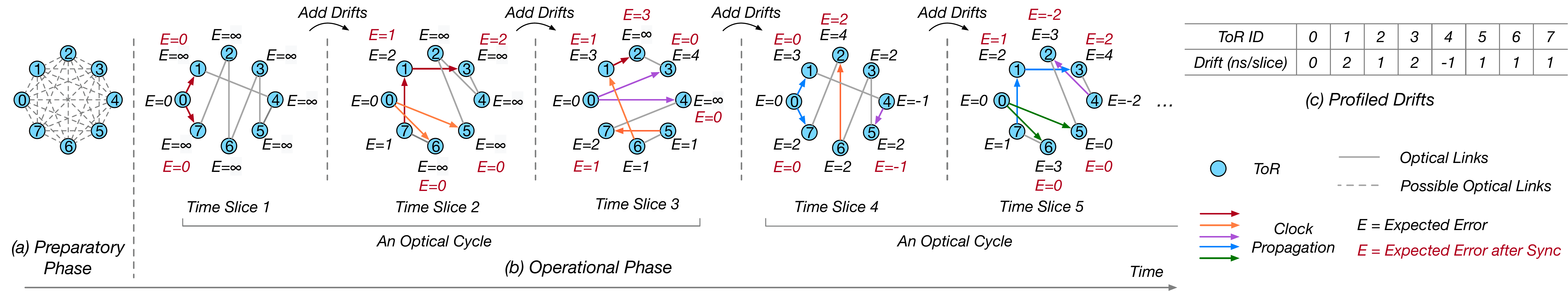}
    \figcap{A running example of \name. (a) Preparatory phase of \name where the ToR-wise connections rotate for parameter profiling. (b) Construction of multi-trees based on the expected sync errors, a new tree rooted at the master ToR ($ToR_0$) per time slice (in different colors). (c) Profiled median value of drifts per ToR for the calculation of expected sync errors in (b).}\label{fig:tree-example}
\end{figure*}

Given the profiled drifts and the fixed optical schedule of circuit connections, the sync planner computes the clock propagation trees over the changing circuits. Sync message exchange in Fig.~\ref{fig:ptp} serves as a building block for every hop of synchronization on the trees. The dynamic circuits produce not a single tree, but \textit{multi-trees}, where each ToR can sync with different parents across time slices.
These trees form the sync plan, specifying which ToR should sync with which other ToR in which time slice. The sync plan is computed in the preparatory phase but executed in the operational phase.
Here, we explain the tree-building process with the example in Fig.~\ref{fig:tree-example}, and the detailed algorithm Alg.~\ref{alg:sync_plan} is in Appx.~\ref{sec:alg}.

The produced trees are illustrated in Fig.~\ref{fig:tree-example}b, with the profiled drifts in Fig.~\ref{fig:tree-example}c. The expected sync error of the master ToR, $ToR_0$, is always $E_0 = 0$.
Before synchronization, the expected sync error for every other ToR, $ToR_i$, is $E_i = \infty$.
Clock propagates from $ToR_i$ to $ToR_j$ if they are connected and $E_i<E_j$, after which $E_j$ inherits $E_i$, i.e., $E_j = E_i$. For instance, in time slice 1 ($t_1$), $ToR_0$ syncs $ToR_1$ and $ToR_7$, setting $E_1 = E_7 = E_0 = 0$. In the next time slice ($t_2$), the profiled drifts are applied, resulting in $E_1 = 0 + 2 = 2$ and $E_7 = 0 + 1 = 1$.

Synced ToRs can then sync others. For example, $ToR_1$ synced by $ToR_0$ in $t_1$ and then sync $ToR_3$ in $t_2$, since $E_1 < E_3$.
Each ToR connects to a set of other ToRs per time slice, but only the one with the lowest expected error becomes the sync parent. For example, in $t_3$, $ToR_2$ syncs with $ToR_1$ rather than $ToR_3$ because $E_1 < E_3$.
If a ToR is not synced in a time slice---because the expected errors of all potential parents are no better---its error only increases by the drift amount. For instance, $ToR_5$ progresses from $E_5 = 1$ in $t_3$ to $E_5 = 2$ in $t_4$.

This procedure continues. Since $E_0 = 0$, the master $ToR_0$ always syncs connected ToRs every time slice, growing a new tree rooted from it,
(shown in different colors in Fig.~\ref{fig:tree-example}b). Each tree will eventually be replaced by newer trees as the errors accumulate over time. In this example, the red tree is replaced by the orange tree (its last leaf $ToR_2$ synced by $ToR_6$ in $t_4$), and the orange tree is replaced by the purple tree (its last leaf $ToR_2$ synced by $ToR_4$ in $t_5$), as the expected errors of the leaf ToRs grow.

The \textit{multi-trees} in \name outperforms the static spanning tree in traditional \dcns (Fig.~\ref{fig:dcn_arch}a) in two ways. (1) We use expected errors to optimize the sync accuracy, unlike the spanning tree that strictly aligns with the network topology and is agnostic to sync errors. (2) The dynamic circuits in optical \dcns pose challenges but also bring opportunities. More ToRs have the chance to sync with the root node or nodes at a shallower depth compared to a static topology.

We do not pass on a clock over multiple ToRs in a time slice,
because the delay of too many sync message exchanges across ToRs may exceed the time slice duration, a challenge that motivates our design of single-message sync ($\S$\ref{sec:single-msg}).

\subsection{Initial Clock and Sync Plan Distribution}\label{sec:distribution}

As motivated in $\S$\ref{sec:intro}, the ToRs need initial clocks of coarse accuracy to identify time slices for the operational phase. 
We reconfigure the optical circuits to form a connected graph and run PTP on this static topology to sync the ToRs with the master ToR, which serves as an intermediary for the optical controller. 
For generality and simplicity, we connect the ToRs as a random graph~\cite{Jellyfish} through circuit reconfiguration, randomly connecting each ToR port to a port on a different ToR. We run connectivity check to guarantee the created topology is fully connected and not partitioned.

The sync plan for the first $M$ optical cycles is distributed to the ToRs over the static network. As the algorithm continues, only sync plan updates (deltas) for later cycles are sent to the ToRs over the changing circuits during the operational phase. In most cases, we observe zero deltas, as the sync plan quickly converges after a few cycles due to the repetitive optical schedule and fixed drifts. As shown in Table~\ref{tb:hardware_usage}, storing sync plans on ToRs requires minimal hardware resources.

We present each sync action in the sync plan as a key-value pair, e.g., $<t, ToR>$ for syncing with a specific $ToR$ in time slice $t$, and both the sync parent and sync child need the sync plan for failure tracking ($\S$\ref{sec:failure}).
We show in Table~\ref{tb:overehad} of Appx.~\ref{sec:appx_overhead} that the total data size for the sync plan
is on the kB scale for a large-scale optical \dcn with 192 ToRs, which has negligible usage of the circuit bandwidth.

%% file: design_operation.tex
\section{\name Operational Phase}\label{sec:operational}

In this section, we show enforcement of the sync plan on ToRs ($\S$\ref{sec:enforcement}), as well as 
single-message sync ($\S$\ref{sec:single-msg}) and timestamp correction ($\S$\ref{sec:hop-error-comp}) as two optimization methods to improve sync efficiency and accuracy.
We further introduce how \name handles different types of failures ($\S$\ref{sec:failure}).

\subsection{Sync Plan Enforcement}\label{sec:enforcement}

In the operational phase, the ToRs enforce the sync plan by loading the received sync actions ($\S$\ref{sec:distribution}) from the switch control plane to the lookup table on the data plane. The sync actions per time slice translate into one combined lookup entry, i.e., $<t, [ToR_1, ToR_2, ..., ToR_n]>$ for syncing with a number of $ToRs$ in time slice $t$. 

We leverage the on-chip packet generator~\cite{joshi2019timertasks} offered by programmable switches to trigger sync messages. Packet generator sends packets into the ingress pipeline reliably according to the configured time interval. In \name, we configure the time interval of the packet generator as the time slice duration, so that on the sync parent ToR, $n$ pre-defined sync messages are generated per time slice for the $n$ possible sync children.
These messages act as the first sync message (out of three) in Fig.~\ref{fig:ptp}. They have the destination fields filled by checking the lookup table and are sent to the children ToRs. The ToR stores the clock offset (Eqn.~2 in Fig.~\ref{fig:ptp}) after the sync action.
The packet generator is configured according to the clock offset to stay in sync with the optical schedule. We defer this process to Appx.~\ref{sec:pkg-gen-sync}.

We perform outlier detection to filter out inaccurate syncs, possibly caused by heavy/bursty traffic, jumbo packets, and other runtime errors. Each ToR maintains the sliding average of the clock offsets, and a new offset beyond the average $\pm$ the statistical tail sync error, e.g., 28ns in our evaluated network ($\S$\ref{sec:eval}), is considered an outlier. It is ignored and the ToR keeps the old offset value. The ToR will get accurate syncs in later time slices from different sync parents. 

According to Table~\ref{tb:overehad} in Appx.~\ref{sec:appx_overhead}, for a large-scale 192-ToR optical \dcn, each ToR has up to 177 entries in the lookup table and takes up to 2.8KB switch memory, which is well under the capacity limit of commercial switches~\cite{miao2017silkroad}.

\subsection{Single-Message Synchronization}\label{sec:single-msg}

Recall from Fig.~\ref{fig:ptp} that each sync operation in traditional sync protocols requires three sync messages, one for clock propagation and two for measuring the propagation delay. The exchange of these three messages takes three (one-way) propagation delays, along with two processing delays taking at least 1$\mu$s each~\cite{DPTP}.
The cable length in production \dcns varies between several to hundreds of meters~\cite{guo2016rdma}. With the propagation delay of 5ns/m~\cite{FiberLatency}, the sync process over a 300m cable takes at least 6.5$\mu$s,
exceeding the time slice duration of many optical \dcns.

To ensure the success of the sync action, the exchange of the three sync messages must finish within one time slice, otherwise the optical circuit would be gone.
The remaining messages can either wait until the next optical cycle when the circuit reappears or be routed through other ToRs where the circuit is available.
However, in the former case, the two successive messages have a time gap of one cycle, incurring large drifts and sync errors. For the latter, the messages traverse paths of different lengths, violating the basic assumption of path symmetry in the delay measurement shown in Fig.~\ref{fig:ptp}.

We solve this problem with single-message sync, where we only keep the initial sync message for clock propagation and eliminate the following two for real-time propagation delay measurement. The propagation delays are stable and profiled in the preparatory phase as described in $\S$\ref{sec:profiling}.
This change only reduces the number of sync messages, without changing the data size of the sync plan or the number of lookup table entries.

\subsection{Timestamp Correction}\label{sec:hop-error-comp}

As shown in Fig.~\ref{fig:ptp}, the sync process involves TX and RX timestamps of the sync parent and child ToRs. Commercial switches support precise timestamping with nanosecond resolution~\cite{aristaPTP,geant2022P4Timestamping}, so they provide accurate RX timestamps capturing the arrival time of packets. It is challenging, though, to obtain accurate TX timestamps, because the actual TX timestamp of a packet is only attainable \emph{after} the packet has been transmitted by the TX MAC. One solution is two-step PTP that sends a follow-up packet to convey the TX timestamp~\cite{ptp}. However, as confirmed by a recent implementation~\cite{DPTP}, the TX timestamp of a sent packet can only be retrieved later from the control plane. 
This incurs latency of at least hundreds of microseconds, making it infeasible for the short time slices of optical \dcns.

The latest one-step PTP has been supported by some switches
~\cite{DPTP, cisco_nexus9000}, which takes the timestamp \emph{before} a packet is passed to the TX MAC and writes it into the packet~\cite{ptp}.
The common practice is to take the timestamp when the packet enters the egress pipeline and add the egress delay to approximate the actual TX timestamp. 
Depending on the port layout of the switch architecture, the MAC clock of each port lags behind the switch clock by different amounts due to varying distances in clock propagation, which results in port-specific errors in TX timestamps.

We correct the TX timestamp errors with a mathematical approach. We model the TX timestamp error as $e$, so the actual TX timestamp is $T = T' - e$, where $T'$ is the measured TX timestamp. Similarly, in the following derivation, we use symbols with primes to denote measured values (with errors) and symbols without primes to denote ground-truth values. Among the four timestamps in Fig.~\ref{fig:ptp}, $T_{1}^{A}$ and $T_{2}^{B}$ are TX timestamps, and $T_{2}^{A}$ and $T_{1}^{B}$ are RX timestamps. Because RX timestamps are accurate while TX timestamps contain errors, we can rewrite Eqn.~1 and Eqn.~2 in Fig.~\ref{fig:ptp} by considering the TX timestamp errors, i.e., $T = T' - e$.

\vspace{-10pt}
{\footnotesize
\begin{equation}\tag{$3$}\label{eqn:delay_ab}
\begin{split}
delay'_{AB} &= \frac{T_{2}^{A} - T_{1}^{A'}}{2} - \frac{T_{2}^{B'} - T_{1}^{B}}{2}  \\[-3pt]
&=\frac{T_{2}^{A} - (T_{1}^{A} - e^{A})}{2} - \frac{(T_{2}^{B} - e^{B}) - T_{1}^{B}}{2}   \\[-3pt]
&= \frac{T_{2}^{A} - T_{1}^{A}}{2} - \frac{T_{2}^{B} - T_{1}^{B}}{2} + \frac{e^{A}+e^{B}}{2} \\[-3pt]
&= delay_{AB} + \frac{e^{A}+e^{B}}{2}
\end{split}
\end{equation}
}

\vspace{-0.10in}
{\footnotesize
\begin{equation}\tag{$4$}\label{eqn:offset_ab}
\begin{split}
offset'_{AB} &= T_{1}^{B} - delay'_{AB} - T_{1}^{A'} \\[-4pt]
&= T_{1}^{B} - (delay_{AB} + \frac{e^{A}+e^{B}}{2}) - (T_{1}^{A} - e^{A}) \\[-4pt]
&= (T_{1}^{B} - T_{1}^{A} - delay_{AB}) + e^{A} - \frac{e^{A}+e^{B}}{2} \\[-4pt]
&= offset_{AB} + \frac{e^{A}-e^{B}}{2}
\end{split}
\end{equation}
}

In Eqn.~\ref{eqn:offset_ab}, we denote the difference between the measured and true clock offsets as
$\Delta_{AB} = (e^{A}-e^{B})/2$,
and we eliminate that term by introducing a reference ToR C. For simplicity, here we present the case where ToR C has the same propagation delay (or same cable length) with ToR~A and ToR~B, i.e., $delay_{AC} = delay_{BC}$.
The general case for any arbitrary reference ToRs is shown in Appx.~\ref{sec:correction_supp}.
If $delay_{AC} = delay_{BC}$, reusing Eqn.~\ref{eqn:delay_ab} on $delay_{AC}$ and $delay_{BC}$, we have Eqn.~\ref{eqn:torc}.

\vspace{-10pt}
{\footnotesize
\begin{equation}\tag{$5$}\label{eqn:torc}
\begin{split}
delay_{AC} &= delay_{BC}\\[-4pt]
delay'_{AC} &= delay_{AC} + \frac{e^{A}+e^{C}}{2} \\[-4pt]
delay'_{BC} &= delay_{BC} + \frac{e^{B}+e^{C}}{2} \\[-4pt]
\end{split}
\end{equation}
}

Combining the equations in Eqn.~\ref{eqn:torc}, we derive $\Delta_{AB}$ in Eqn.~\ref{eqn:final}, indicating $\Delta_{AB}$ can be obtained by measuring the propagation delays between ToRs A and C and between ToRs B and C.

\begin{figure}[tbp]
    \centering
    \includegraphics[width=\linewidth]{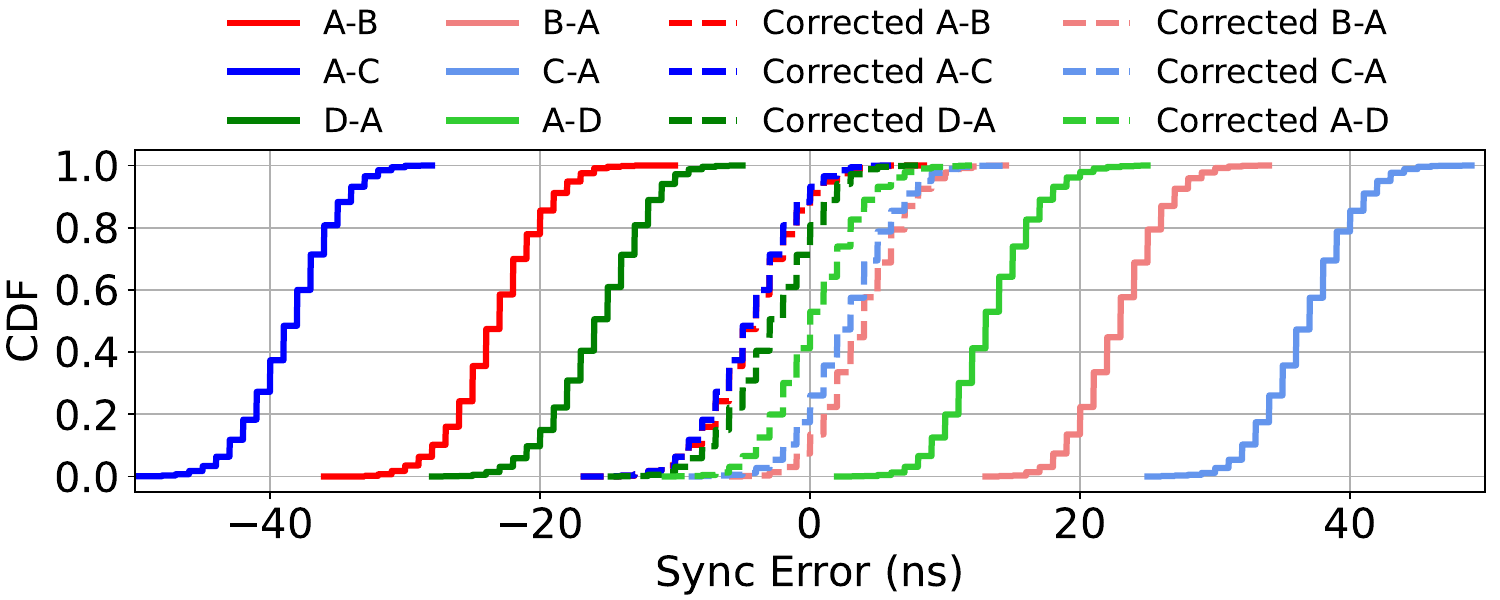}
    \figcap{Timestamp correction for different port pairs.}\label{fig:sync_system_error}
\end{figure}

\vspace{-10pt}
{\footnotesize
\begin{equation}\tag{$6$}\label{eqn:final}
\begin{split}
delay'_{AC} - delay'_{BC} = \frac{e^{A}-e^{B}}{2} = \Delta_{AB}
\end{split}
\end{equation}
}

We validate this method in Fig.~\ref{fig:sync_system_error}, where we sync different port pairs of an Intel Tofino2 switch with one-step PTP.
The ports A--D on the same switch share the same physical clock,
so the ground-true clock offset after synced by another port ($offset_{AB}$ in Eqn.~\ref{eqn:offset_ab}) is always zero. Taking that into Eqn.~\ref{eqn:offset_ab}, we have $offset'_{AB} = (e^{A}-e^{B})/2$, showing the measured clock offset (or sync error in Fig.~\ref{fig:sync_system_error}) reflects the ports' TX timestamp errors.
That explains why the port pairs in Fig.~\ref{fig:sync_system_error} exhibit different sync errors (or offsets), and swapping the sync parent and child ports reverses the error distributions. These results motivate the need for timestamp correction. In Fig.~\ref{fig:sync_system_error}, without correction, the median of relative timestamp errors between ports can be up to 38ns, while our approach reduces the median of timestamp errors to within 4ns.

\begin{figure}[tbp]
    \centering
    \includegraphics[width=\linewidth]{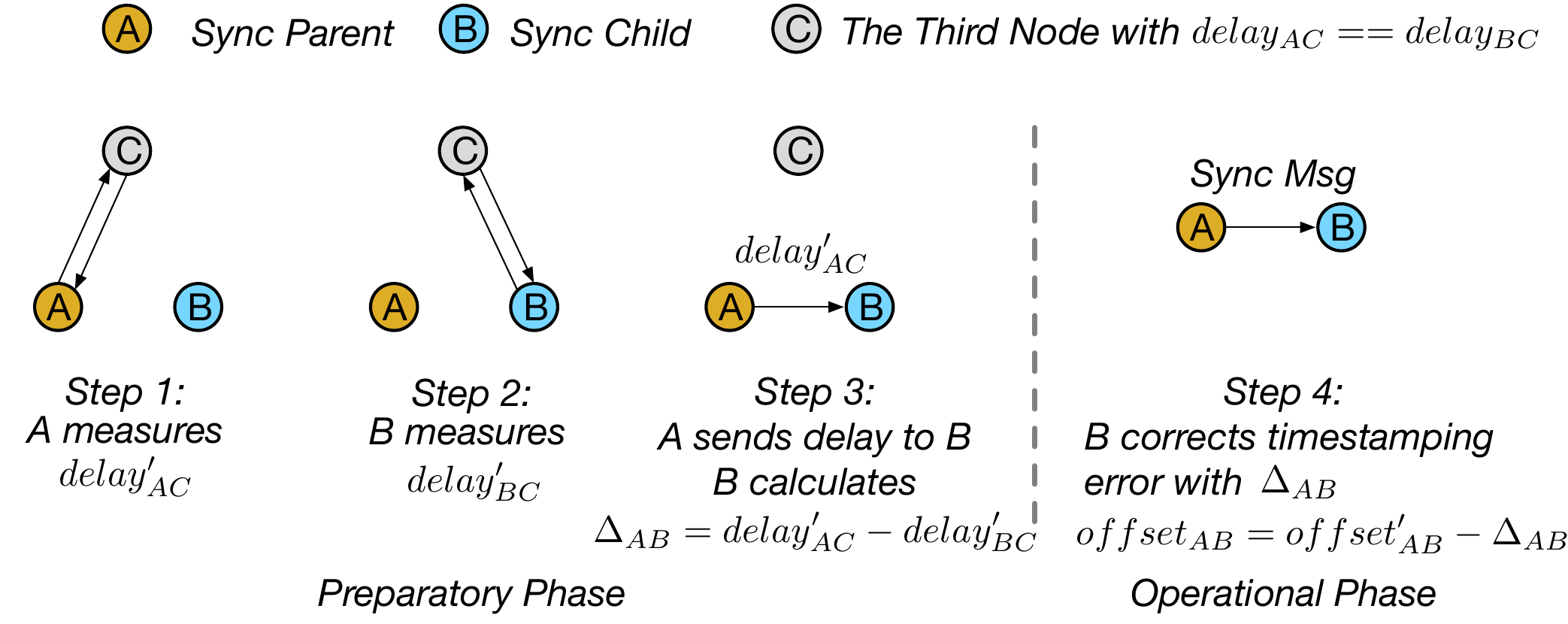}
    \figcap{TX timestamp error correction based on Eqn.~\ref{eqn:delay_ab} -- \ref{eqn:final}.}\label{fig:compensation}
\end{figure}

The specific error correction process in \name is illustrated in Fig.~\ref{fig:compensation}. Given the cabling plan of the \dcn, we find a reference ToR C for every pair of ToR A and ToR B, where AC and BC have the same cable length. As explained in $\S$\ref{sec:profiling}, propagation delays of every pair of ToR ports are profiled in the preparatory phase, including $delay'_{AC}$ and $delay'_{BC}$ (\textit{Steps 1, 2}). After that, the ToRs exchange these delays, so that a child ToR, such as B in Fig.~\ref{fig:compensation}, can calculate $\Delta_{AB}$ regarding every potential parent ToR A (\textit{Step 3}). Then in the operational phase, the child ToR can apply the stored $\Delta_{AB}$ to get the accurate clock offset relative to a sync parent (\textit{Step 4}).

In case no reference ToR C with the same cable length to ToRs A and B can be found, we choose an arbitrary reference ToR C and modify Eqn.~\ref{eqn:delay_ab} -- \ref{eqn:final} to account for the cable length difference between AC and BC in the calculation of $\Delta_{AB}$. This generalized method is described in Appx.~\ref{sec:correction_supp}.

\subsection{Failure Handling}\label{sec:failure}

\name is robust to failures by nature. Even without explicit failure handling, the multi-trees in \name
allow ToRs to sync with different parents in different time slices.
We further strengthen fault tolerance by tackling the following failures. We assume a failure-assured optical controller offered by the optical \dcn fabric.

\para{Sync planner and master ToR failures.}
We deploy a backup sync planner following the primary-backup model. The primary sync planner sends keepalive messages to the backup one and fails over to it when lost.
We also deploy a backup master ToR, which acts as a hot standby of the primary. The sync planner computes clock propagation trees rooted at both the primary and backup master ToRs, and the optical controller syncs both of them,
but
they are not part of each other's trees to avoid multiple clock sources.

\para{Link failures.}
We make the sync algorithm ($\S$\ref{sec:sync-plan}) compute backup sync parents with the second smallest estimated sync errors.
If the parent is the master ToR, the backup parent is the backup master ToR. 
A child ToR can detect a link failure if it has not been synchronized over an expected circuit for two consecutive times, and it notifies the backup parent ToR to push the clock towards it. 
It also reports the failure to the sync planner\footnote{The optical schedule of an optical \dcn guarantees that every ToR is connected to every other ToR at least once in an optical cycle, so under a small number of link or ToR failures, a child ToR has high probability to reach the sync planner through optical routing on the optical schedule~\cite{Opera, RotorNet}.}, which removes the links and computes the global optimal sync plan for future cycles.

\para{ToR and OCS failures.}
A ToR failure can be regarded as concurrent link failures of all the ToR's optical ports. As introduced in $\S$\ref{sec:background}, an optical \dcn fabric consists of a number of OCSes. An OCS may experience partial failures, e.g., a proportion of circuits down, or some circuits still connected but not being able to be reconfigured. The worst case of a completely down OCS causes one link disconnected per ToR (Fig.~\ref{fig:dcn_arch}b). All these cases can be viewed as link failures.

\para{Cooling failures.}
This case happens when the \dcn cooling malfunctions or switches overheat, e.g., caused by high traffic load~\cite{Sundial}. 
The temperature increase changes drifts and makes our trees built from the profiled drifts invalid.
ToRs can detect cooling failures with the equipped temperature sensors~\cite{graham} and notify the sync planner,
which then re-profiles the drifts and updates the trees
live while the \dcn is continuously operating ($\S$\ref{sec:profiling}).

%% file: testbed.tex
\section{Prototype Testbed}\label{sec:testbed}

\begin{figure}[tbp]
    \centering
    \begin{subfigure}[t]{0.49\columnwidth}
        \includegraphics[width=\linewidth]{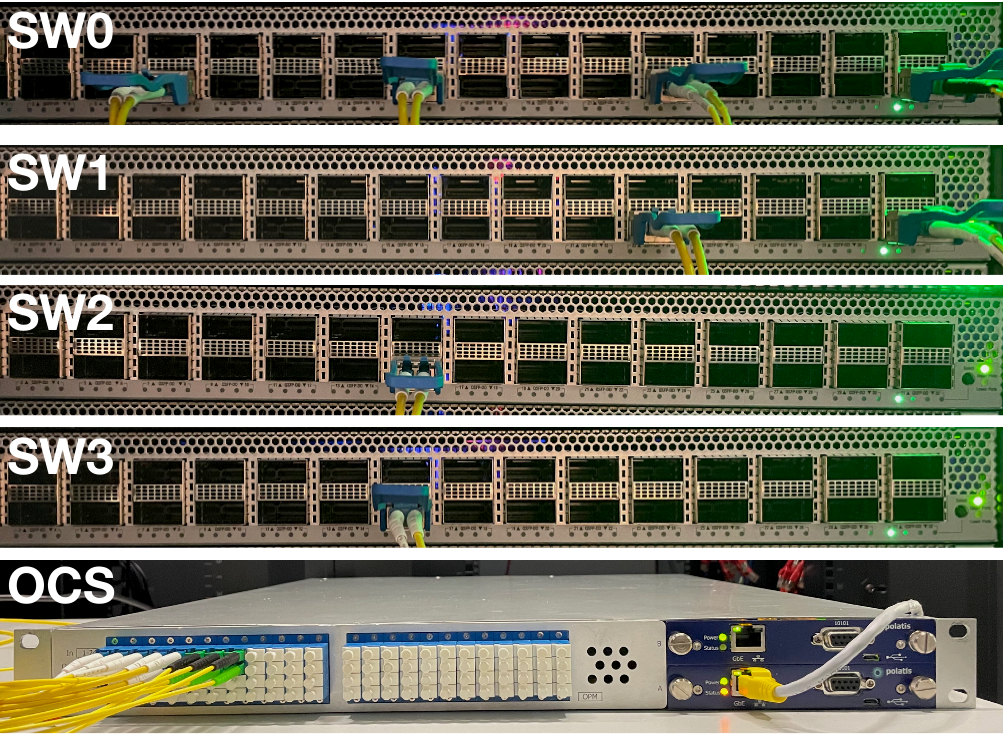}
        \label{fig:testbed_diagram}
    \end{subfigure}
    \hfill
    \begin{subfigure}[t]{0.48\columnwidth}
        \includegraphics[width=\linewidth]{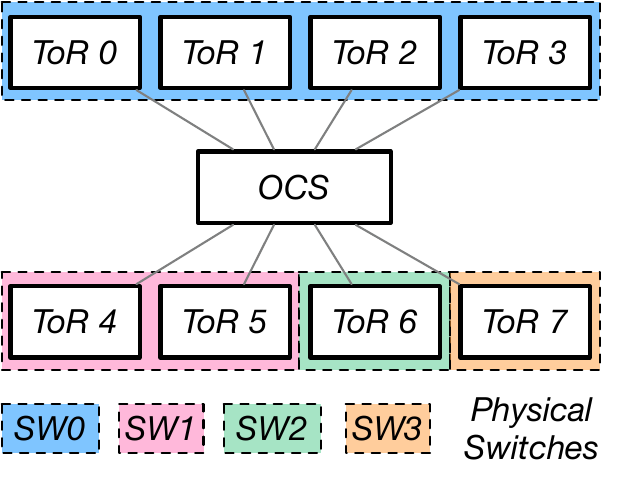}
        \label{fig:testbed_pic}
    \end{subfigure}
    \figcap{Setup of the optical testbed.}\label{fig:testbed}
\end{figure}

We implement \name on programmable switches and build an optical testbed to validate our implementation. Large-scale evaluation of \name is in $\S$\ref{sec:eval}.

As Fig.~\ref{fig:testbed} shows, our testbed consists of a 24$\times$24 MEMS \ocs and 4 EdgeCore DCS810 Intel Tofino2 switches, where 8 logical ToRs are realized on the physical switches (colored in the figure) and each connected to the \ocs through a 3m long fiber at 100Gbps. $ToR_0$ is the master ToR for synchronization. Although nanosecond-speed \ocses \name targets at are technically mature (Fig.~\ref{fig:ocs_hardware}), they are expensive nowadays due to the lack of mass production\footnote{Commercial nanosecond-speed \ocses have a small number of ports~\cite{ns_OCS_epi, ns_OCS_agiltron}. For example, 8$\times$8 connectivity required in our testbed can be realized with 6 4$\times$4 nanosecond-speed \ocses, costing $\sim$\$120k in total~\cite{ocs_price}.}. So, we use a more affordable MEMS-based \ocs with 10ms reconfiguration delay and set the time slice duration to 1s. MEMS has the same behaviors as nanosecond-speed \ocses from the system's perspective. The only difference is longer time slices, which causes higher drifts and serves as a stress test for \name.

\paralit{\textbf{Correctness of drift profiling.}}
We first validate our drift profiling process, the prerequisite for drift-aware synchronization. As described in $\S$\ref{sec:profiling}, in the preparatory phase, we reconfigure the \ocs to connect each ToR to the master ToR in turn. The collected drifts from the three physical switches $SW_1$-$SW_3$ relative to $SW_0$, where $ToR_0$ resides, exhibit stable medians across various measurement intervals, with a variance range of 24ppm (specific numbers in Fig.~\ref{fig:drift_measurement} of Appx.~\ref{sec:appx_drift}). We also re-profile the drifts in the operational phase, when the \ocs follows a round-robin schedule, and the results are similar. 
Our observation aligns with the drift characteristics presented in Graham~\cite{graham}, thereby confirming the correctness of our method.

\paralit{\textbf{Effectiveness of tree building.}}
Next, we validate the drift-aware trees. As illustrated in Fig.~\ref{fig:testbed}, we make $ToR_1$-$ToR_3$ observed ToRs and co-locate them with $ToR_0$ on $SW_0$ to measure their sync errors. The ground-truth sync offset between co-located ToRs is zero, so the offset produced by \name is the sync error.
As $ToR_1$-$ToR_3$ have no drift relative to $ToR_0$, we add profiled drifts from $SW_1$-$SW_3$ to them, such that they are at maximum 1-3 hops from the master ToR in the trees.

\begin{figure}[tbp]
    \centering
    \includegraphics[width=1\columnwidth]{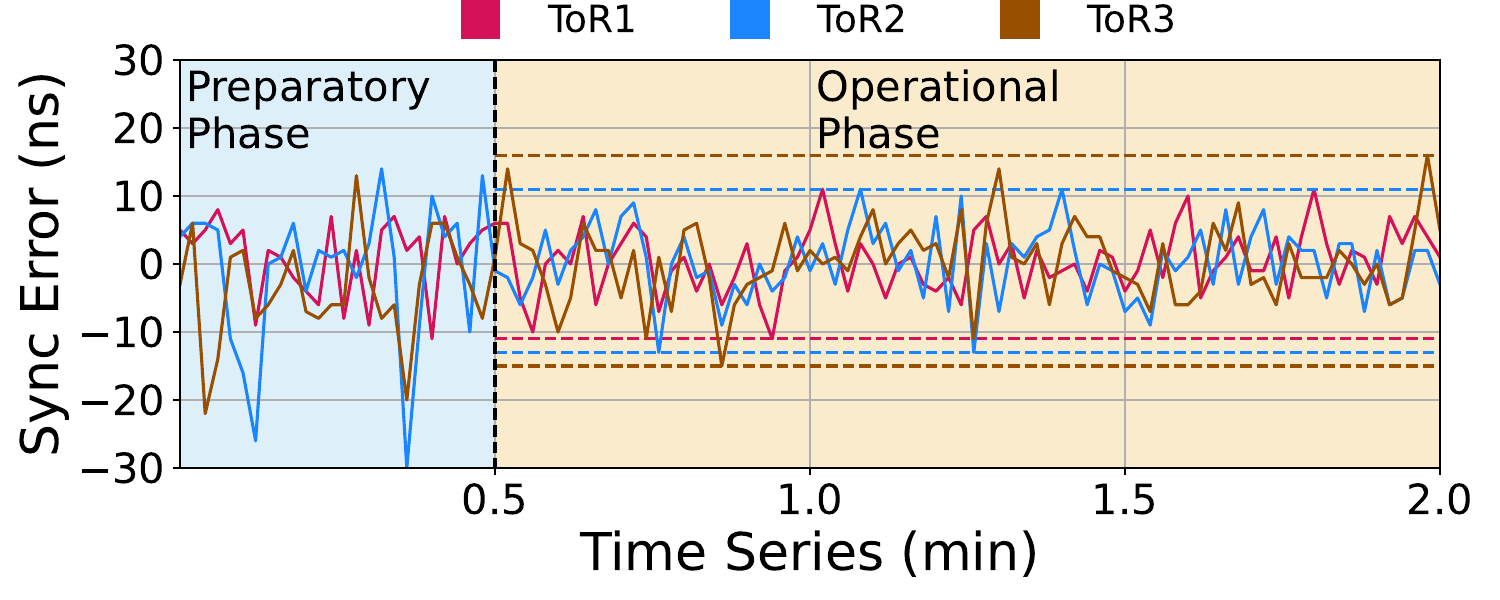}
    \figcap{Sync errors of $ToR_1$-$ToR_3$ on the testbed (Fig.~\ref{fig:testbed}), which are at maximum 1, 2, 3 hops from the master ToR $ToR_0$ throughout the sync process. 
    }
    \label{fig:testbed_three_sw}
\end{figure}

Fig.~\ref{fig:testbed_three_sw} plots the development of sync errors over time. The preparatory phase has sync errors under 30ns, and the operational phases further pushes them under 16ns.
$ToR_1$-$ToR_3$ have different depths in the clock propagation trees thus having different sync error ranges.
The max hop count is the worst case in the ToR's entire sync process, e.g., max 1 hop is always syncing with the master ToR, and max 2 and 3 hops are for syncing with other parent ToRs sometimes.
ToRs with more sync hops do exhibit higher errors, but \name successfully brings down their sync errors over time. This result shows the advantage of multi-trees.
Even if a ToR is deep down a sync tree for one time slice, it has the chance to switch to shallower trees in later time slices.

\begin{table}[tbp]
\centering
\footnotesize
\caption{Hardware resource usage on Tofino2 switches}
\begin{tabular}{lllll}
\hline
       & \begin{tabular}[c]{@{}l@{}}Match\\ Xbar\end{tabular} & SRAM   & \begin{tabular}[c]{@{}l@{}}VLIW\\ Instructions\end{tabular} & \begin{tabular}[c]{@{}l@{}}Register\\ ALU\end{tabular} \\ \hline
PTP    & 3.39\%                                               & 3.33\% & 6.25\%                                                      & 3.12\%                                                 \\
\name & 5.96\%                                               & 5.62\% & 5.21\%                                                      & 2.78\%                                                 \\ \hline
\end{tabular}\label{tb:hardware_usage}
\end{table}

\paralit{\textbf{Hardware resource usage.}}
Finally, we compare the hardware resource usage of \name with DPTP~\cite{DPTP}, a PTP implementation in the programmable switch data plane.
Table~\ref{tb:hardware_usage} presents the hardware usage of a ToR for the large optical \dcn setting with 192 ToRs in $\S$\ref{sec:eval}.
\name is light-weighted, consuming less than 6\% resources of all types on Tofino2 switches. It takes slightly more crossbar and SRAM than PTP, by the match-action table for enforcing sync plans ($\S$\ref{sec:enforcement}). 
Single-message sync ($\S$\ref{sec:single-msg}) simplifies the computation of sync offset, as it uses the profiled propagation delay and spares the runtime calculation of it (Fig.~\ref{fig:ptp}). Thus, \name takes less computation resources, including VLIW instructions and register ALUs.
More detailed overhead analysis is in Appx.~\ref{sec:appx_overhead}, which shows \name needs at most 177 table entries per ToR, generates 33.4KB total data for the sync plan, and consumes 25.8Mbps peak bandwidth for sync messages, all minimal for modern switches.

%% file: eval.tex
\section{Evaluation}\label{sec:eval}%

We have validated the correct functionality of \name in $\S$\ref{sec:testbed} with an \ocs and multiple programmable switches. In this section, we conduct emulation experiments for a 192-ToR optical \dcn to evaluate the performance of \name at scale.

\para{Experimental Setup.}
To analyze the key technique of drift-aware synchronization in \name, we need to generate drifts in a controlled manner and measure sync errors accurately. Thus, similar to $\S$\ref{sec:testbed}, we realize the 192 ToRs as logical ToRs on one Tofino2 switch to make them share the same clock.
Many synchronization proposals follow this standard approach for accurate sync error measurement~\cite{huygens,DPTP}. Other methods, such as comparing hardware-generated pulse-per-second (PPS) signals using an oscilloscope or physical-layer logging~\cite{dtp} require specialized NICs, which is infeasible on commercial switches.
To emulate realistic drifts,
we assign each logical ToR a median drift sampled from a real-world drift distribution~\cite{huygens} with the variance from Graham~\cite{graham}.

We connect two ports of the Tofino2 switch with a 3m long loopback fiber at 100Gbps to emulate all possible circuit connections of the \ocses. This is because \ocses are physical-layer devices and do not add switch hops. They are essentially waveguides like optical fibers. We program the logical ToRs to only be able to send sync messages in the time slices when the circuits are supposedly available. We pace the sync messages to avoid contention, emulating dedicated optical connections between ToRs.
Necessary states per logical ToR, including the profiled drifts and propagation delays ($\S$\ref{sec:profiling}), sync plan ($\S$\ref{sec:sync-plan}), timestamp correction offset ($\S$\ref{sec:hop-error-comp}) and clock offset (Fig.~\ref{fig:ptp}) are stored in the switch registers.

\para{Comparison baselines.}
As the first of its kind work, it is hard to find fair baselines for \name. We compare \name with the \textbf{strawman} solution, where each ToR only synchronizes with the master ToR once per optical cycle. We run \name and the strawman solution on Opera~\cite{Opera}, a well-recognized optical \dcn architecture. Its round-robin optical schedule forms a well-connected \textit{expander} topology every time slice, as illustration in Fig.~\ref{fig:tree-example}b. In the 192-ToR Opera optical \dcn, every ToR connects to 12 \ocses through 12 uplinks, and the time slice duration is  300$\mu$s. We adopt this default setting but also 
vary the time slice duration between 1$\mu$s and 1ms to understand \name's generality.

To put the numbers in context and break down the \name gains, we also compare with sync protocols in traditional \dcns, though they \textbf{do not} work on optical \dcns. 
We choose \textbf{PTP}~\cite{ptp} and \textbf{Sundial}~\cite{Sundial}, because they are the most widely used and most accurate sync solutions for traditional \dcns, respectively, and they also feature tree-building in their designs. We run them on the \textit{Expander} topology~\cite{Jellyfish} that has the closest structure as our optical \dcn but maintains the static topology.
For fair comparison, we configure them to have the same sync frequency as \name.

In the rest of the section, we raise a series of questions to evaluate different aspects of \name.

\para{Q1: What sync accuracy can \name achieve?}

\begin{figure}[tbp]
\begin{minipage}[t]{0.47\linewidth}
    \centering 
    \includegraphics[width=\linewidth]{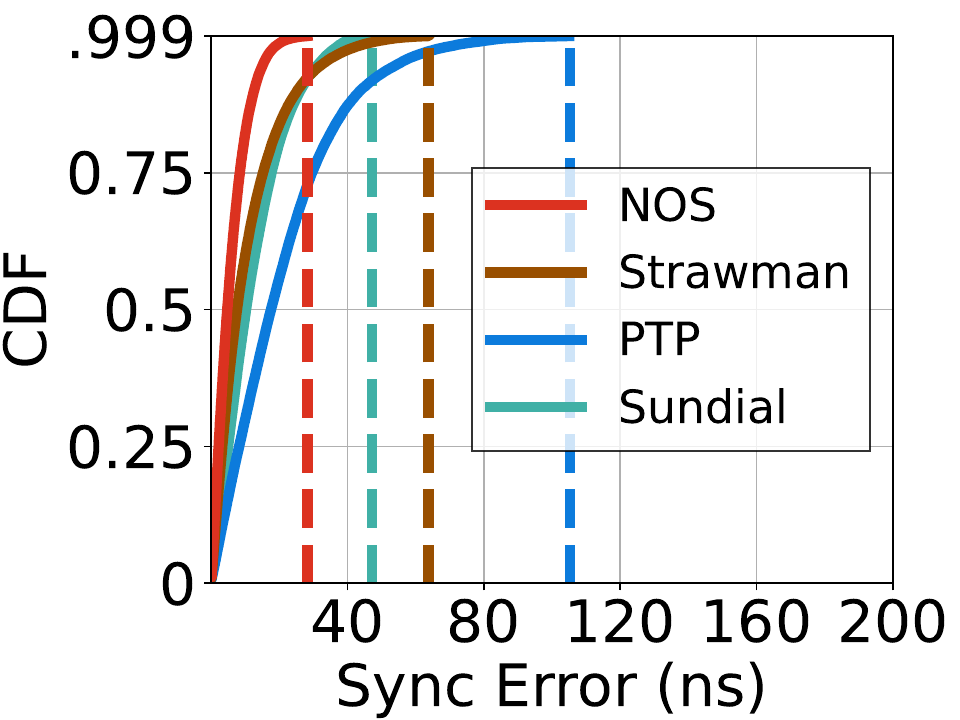}
    \caption{Sync error distributions under 300$\mu$s time slice duration.
    }\label{fig:accuracy-300us}
\end{minipage}
\hspace{0.1in}
\begin{minipage}[t]{0.47\linewidth}
    \centering 
    \includegraphics[width=\linewidth]{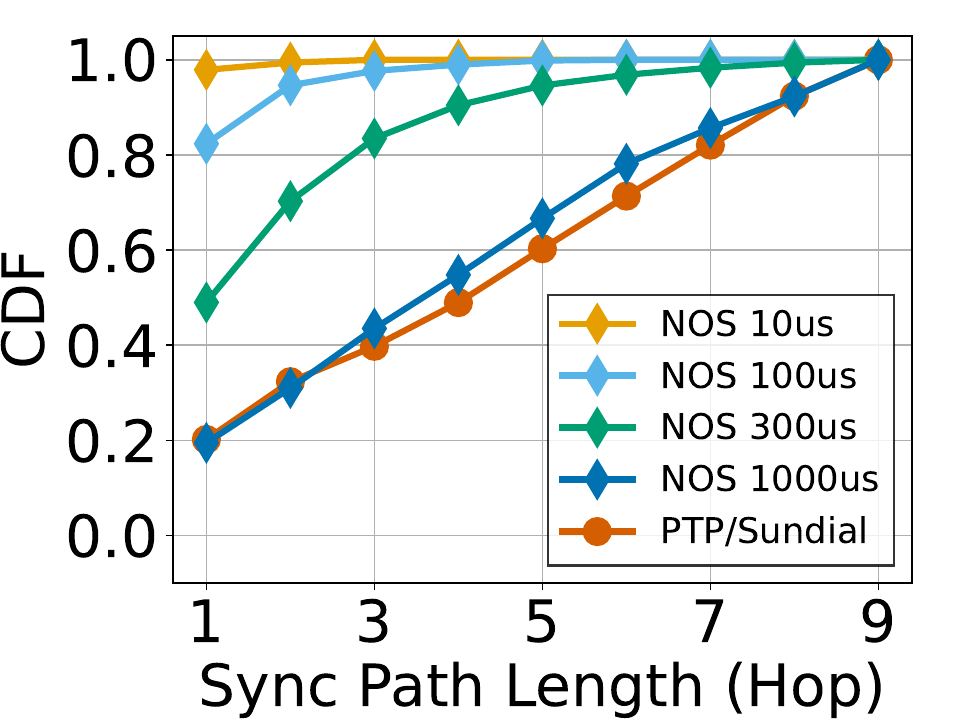}
    \caption{Sync hop count distributions under different slice durations.}
    \label{fig:hop_length_cdf}
\end{minipage}
\end{figure}

\begin{figure*}[tbp]
    \centering
    \begin{minipage}[t]{0.24\linewidth}
        \centering 
        \includegraphics[width=\linewidth]{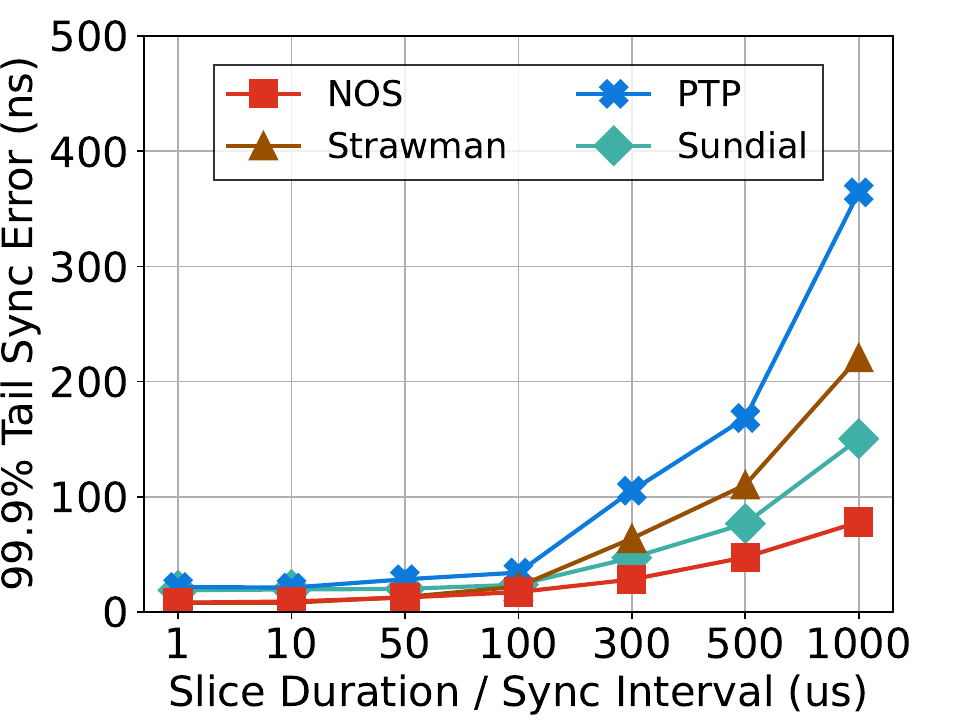}
        \figcap{Tail sync errors under different sync intervals.}\label{fig:sync_interval}
    \end{minipage}
    \hfill
    \begin{minipage}[t]{0.24\linewidth}
        \centering 
        \includegraphics[width=\linewidth]{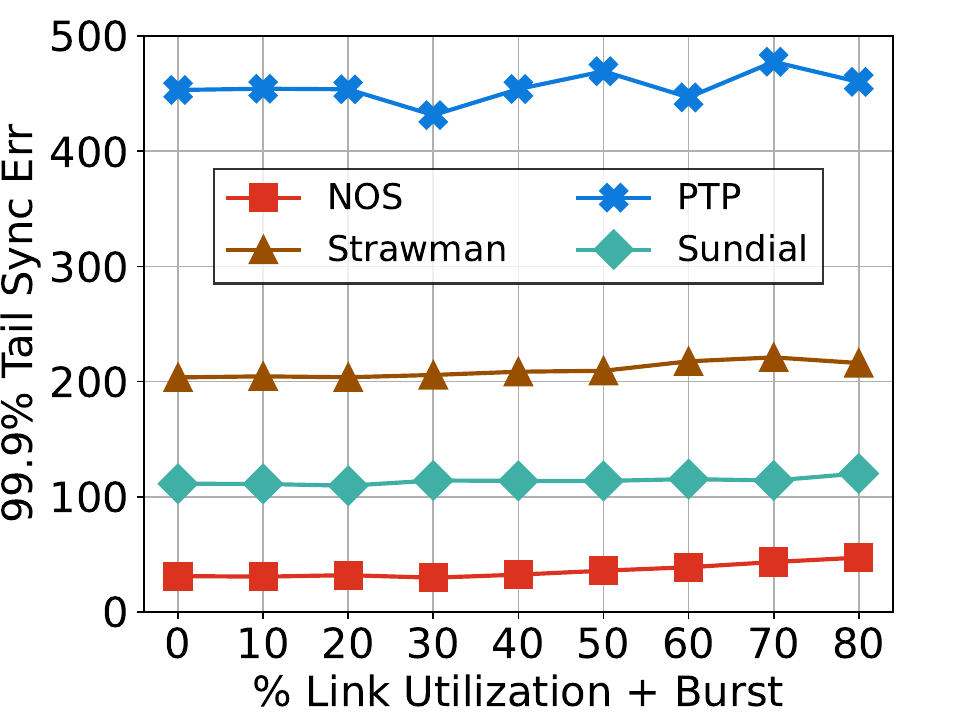}
        \figcap{Tail sync errors under bursty background traffic.}\label{fig:bg_traffic}
    \end{minipage}
    \hfill
    \begin{minipage}[t]{0.24\linewidth}
        \centering 
        \includegraphics[width=\linewidth]{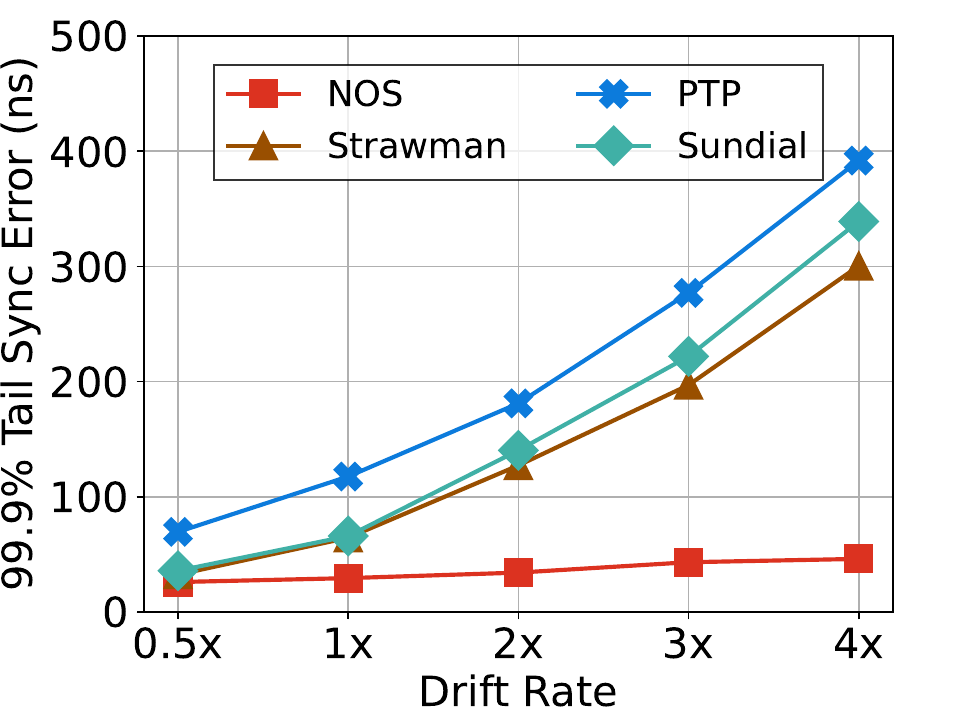}
        \figcap{Tail sync errors under different drift rates.}\label{fig:drift_rate}
    \end{minipage}
    \hfill
    \begin{minipage}[t]{0.24\linewidth}
        \centering 
        \includegraphics[width=\linewidth]{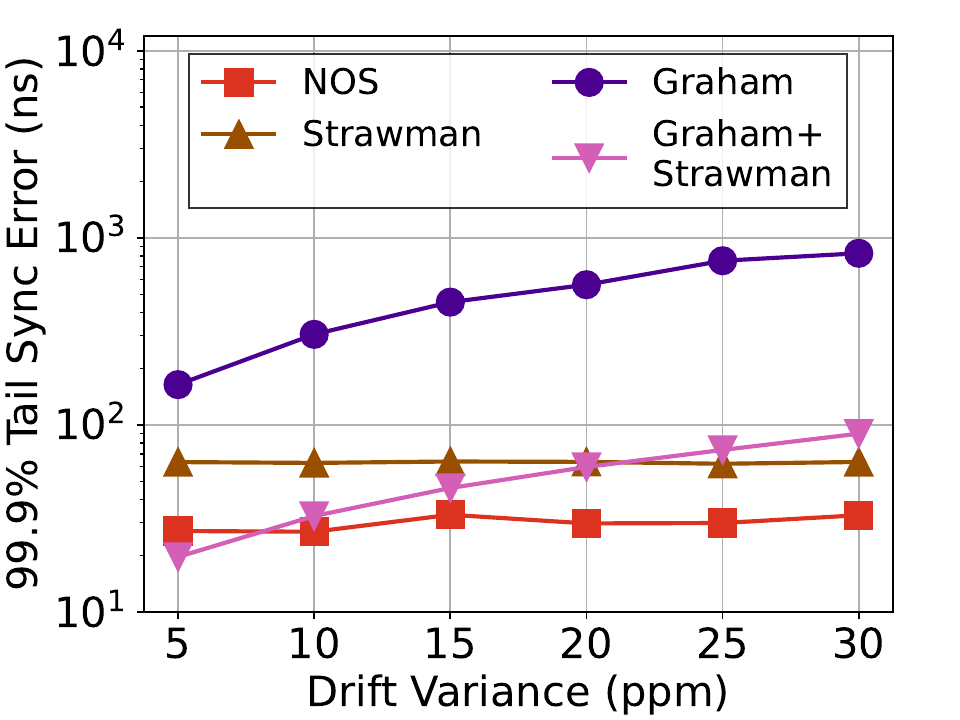}
        \figcap{Tail sync errors under different drift variances.}\label{fig:drift_noise}
    \end{minipage}
\end{figure*}

We first give an overview of the \name performance.
Fig.~\ref{fig:accuracy-300us} presents the sync error distributions of \name and other compared sync solutions under the default setting of 300$\mu$s time slices.
\name achieves 28ns sync accuracy, i.e., 99.9\textsuperscript{th} percentile tail sync error, a 2.3$\times$ improvement to the strawman solution. 
In the strawman solution, a ToR only syncs with the master ToR once per cycle, thus suffering from clock drifts,
while \name solves this problem by allowing synchronization across ToRs and forming more sophisticated clock propagation trees.

The sync accuracy of \name is 3.8$\times$ and 1.7$\times$ better than that of PTP and Sundial, the latest sync protocols for traditional \dcns. Two factors contribute to \name's advantages.
(1) \name builds drift-aware trees guided by the expected sync errors to optimize the sync accuracy, whereas Sundial and PTP use spanning tree that is agnostic to sync errors. (2) The multi-trees in \name give ToRs the chance to sync with different parents with lower sync errors across time, while PTP and Sundial are stuck with the static tree, and one drifted node may influence more children nodes.

We dive into their clock propagation trees to better understand these results. Fig.~\ref{fig:hop_length_cdf} shows the distributions of their sync hop counts for all the sync actions.
\name's drift-aware multi-trees have lower hop counts than the static tree in PTP and Sundial. 
With 300$\mu$s time slices, 90.44\% sync actions in \name are within 4 hops, and 70.28\% within 2 hops. This is because ToRs in \name can sync with different parents across time slices, and they choose parents with minimum expected drifts which tend to be close to the master ToR. Shorter time slices result in even fewer hop counts, as drifts accumulate less and \name builds shallower trees. When the slice duration is long enough, e.g., 1000$\mu$s, one time slice of the optical \dcn is similar to a static network, so the hop counts of \name are comparable to PTP and Sundial. However, as we will show in Fig.~\ref{fig:sync_interval}, the sync accuracy of \name is significantly higher than PTP and Sundial in this case, because \name is drift-aware to choose more accurate sync parents, though at similar hop counts.

\para{Q2: How do different factors affect sync accuracy?}

Next, we analyze different aspects of \name by quantifying the influence of various factors.

\paralit{\textbf{Impact of sync frequency.}}
In Fig.~\ref{fig:sync_interval}, we vary the sync interval of \name by tuning the time slice duration, as each ToR is at most synced once per time slice
($\S$\ref{sec:sync-plan}). 
The strawman solution uses the same time slice durations. Since PTP and Sundial run on static networks and do not have time slices, we configure them to have the same sync intervals as \name. Extending the sync interval increases tail sync errors of all solutions, but \name has the least increase. \name's resistance to longer sync intervals is the result of the drift-aware multi-trees.
Even synced less frequently, \name can still select sync parents with the lowest expected errors, whereas others have to combat higher drifts.

\begin{figure}[tbp]
    \centering 
    \includegraphics[width=0.8\linewidth]{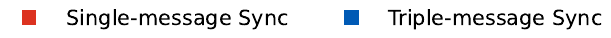}
    \vfil
    \begin{subfigure}[t]{0.49\columnwidth}
        \includegraphics[width=\linewidth]{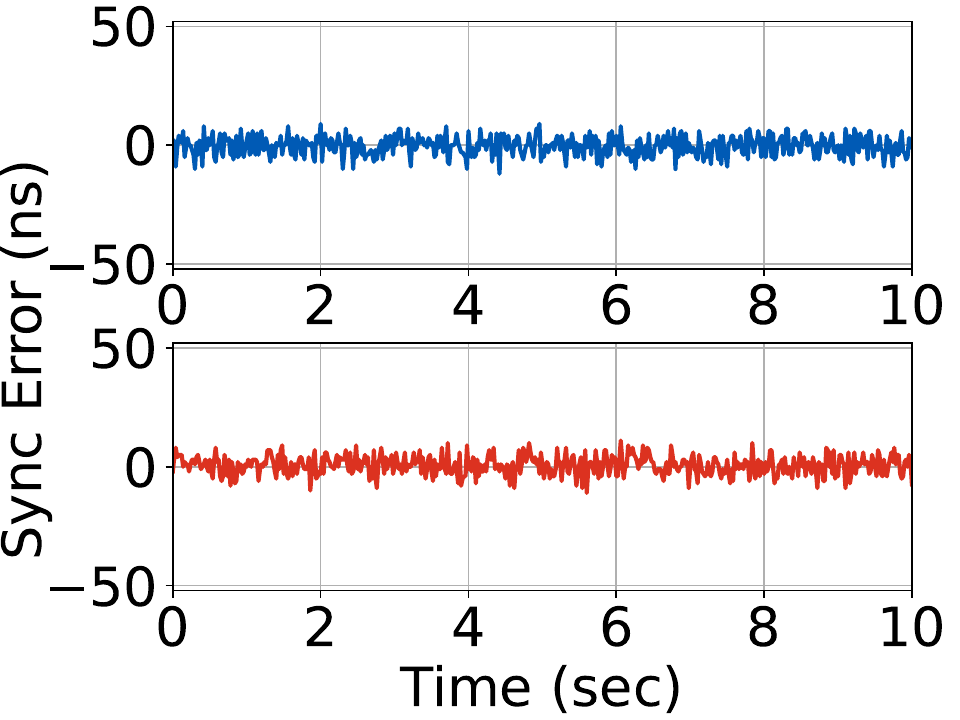}
        \sfigcap{10$\mu$s slice duration
        }\label{fig:sm-sync-10us}
    \end{subfigure}
    \hfill
    \begin{subfigure}[t]{0.49\columnwidth}
        \includegraphics[width=\linewidth]{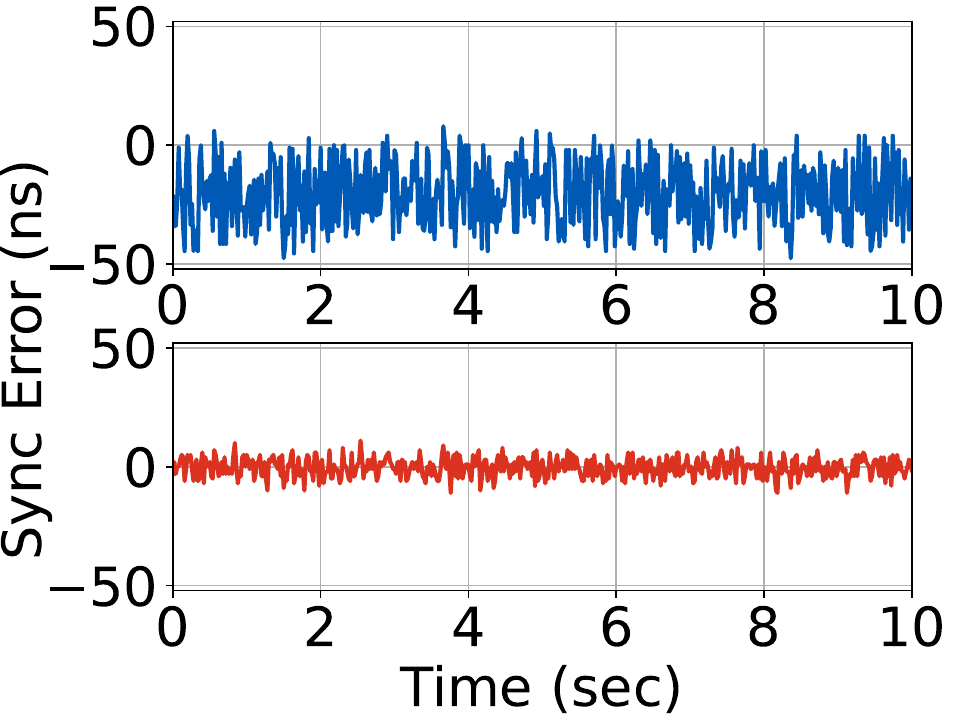}
        \sfigcap{1$\mu$s slice duration
        }\label{fig:sm-sync-1us}
    \end{subfigure}
    \figcap{Time series of errors in triple-message and single-message sync with slice durations of (a) 10$\mu$s and (b) 1 $\mu$s.}
    \label{fig:sm-sync}
\end{figure}

\paralit{\textbf{Impact of background traffic.}}
Data packets, especially large ones, introduce additional egress delays and can cause inaccuracies in the TX timestamp of the subsequent sync message. To evaluate this case, we use a packet generator to create 1500B background traffic from 0 to 80\% link utilization between the connected ports of the the Tofino2 switch.
A measurement study reveals that bursts occur less than 10\% of the time in \dcns~\cite{Bursts}.
As a stress test, we create bursts for 10$\mu$s every 100$\mu$s with 9000B jumbo packets using another packet generator, and we 
follow the common practice to send sync messages 
through prioritized queues~\cite{DPTP, Sundial}.

As Fig.~\ref{fig:bg_traffic} shows, background traffic degrades \name's sync accuracy minimally compared to Fig.~\ref{fig:accuracy-300us}, increasing the tail sync error by only 5ns under the standard 40\% traffic in \dcns and 19ns under the extremely heavy 80\% traffic. 
This is thanks to outlier filtering ($\S$\ref{sec:enforcement}) that skips sync operations whose resulting clock offset is beyond the allowable margin for each parent ToR, and
the effect is similar to increasing the sync interval in Fig.~\ref{fig:sync_interval}. The strawman solution and PTP suffer from inaccurate timestamps and have 3.28$\times$ and 4.47$\times$ degradation of sync accuracy.
Sundial also filters out inaccurate sync operations, but without ToR-to-ToR drift profiling in \name, it has to use a loose margin covering all valid clock offsets network-wide. This margin fails to detect the burst errors, leading to 2.43$\times$ accuracy degradation.

\paralit{\textbf{Impact of drift rate.}}
We vary the drift rate by scaling the distribution of median drift values~\cite{huygens} from 0.5$\times$ up to 4$\times$. From Fig.~\ref{fig:drift_rate}, we observe that \name's tail sync error increases by only 18ns, in contrast to other solutions' increases of hundreds of ns. 
This result demonstrates the ability of \name to tolerate high drifts, by selecting sync parents based on their expected sync errors from the various choices in optical \dcns.

\paralit{\textbf{Impact of drift variance.}}
We further examine varying drift variance, which is not included in our drift profiling and is the cause of estimation errors ($\S$\ref{sec:profiling}). In Fig.~\ref{fig:drift_noise}, we compare the sync accuracy of \name with Graham that compensates drifts locally based on the fitted value~\cite{graham}, with strawman that naively syncs each ToR only with the master ToR, and with these two approaches combined. Graham and Graham+Strawman are sensitive to the range of drift variance, because Graham compensates drifts by the expected value without considering the variance, and compensation alone with Graham results in high sync errors.
Strawman successfully eliminates the effect of drift variance by correcting runtime drifts with the sync process.
\name also overcomes drift variance with active syncing and uses drift-aware trees to reduce sync errors. It is more effective than simply combining compensation and syncing in Graham+Strawman. These results echo the motivating experiment in Fig.~\ref{fig:drift_motivation}, confirming the necessity of drift-aware synchronization.

\begin{figure}[tbp]
    \centering 
    \includegraphics[width=1\columnwidth]{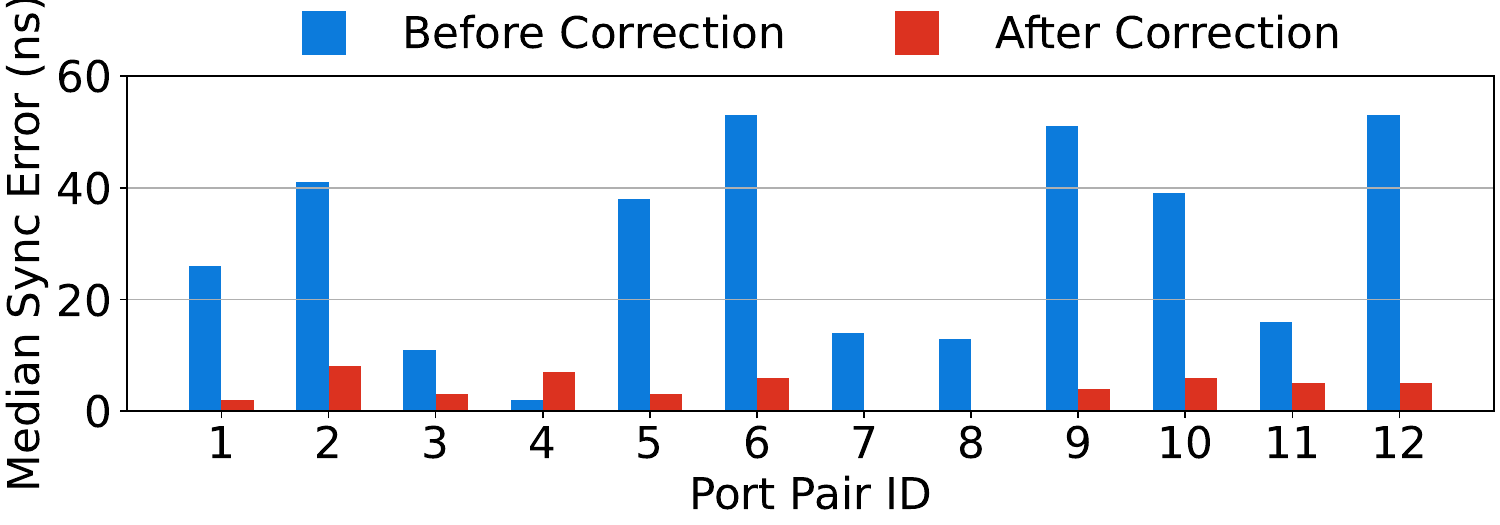}
    \caption{Absolute median sync errors before and after TX timestamp correction for different port pairs.
    }
    \label{fig:compensation_bar}
\end{figure}

\begin{figure*}[h]
    \centering 
    \includegraphics[width=0.99\linewidth]{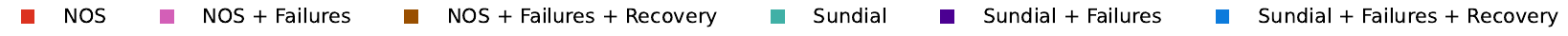}
    \vfil
    \begin{subfigure}{0.245\textwidth}
        \includegraphics[width=\linewidth]{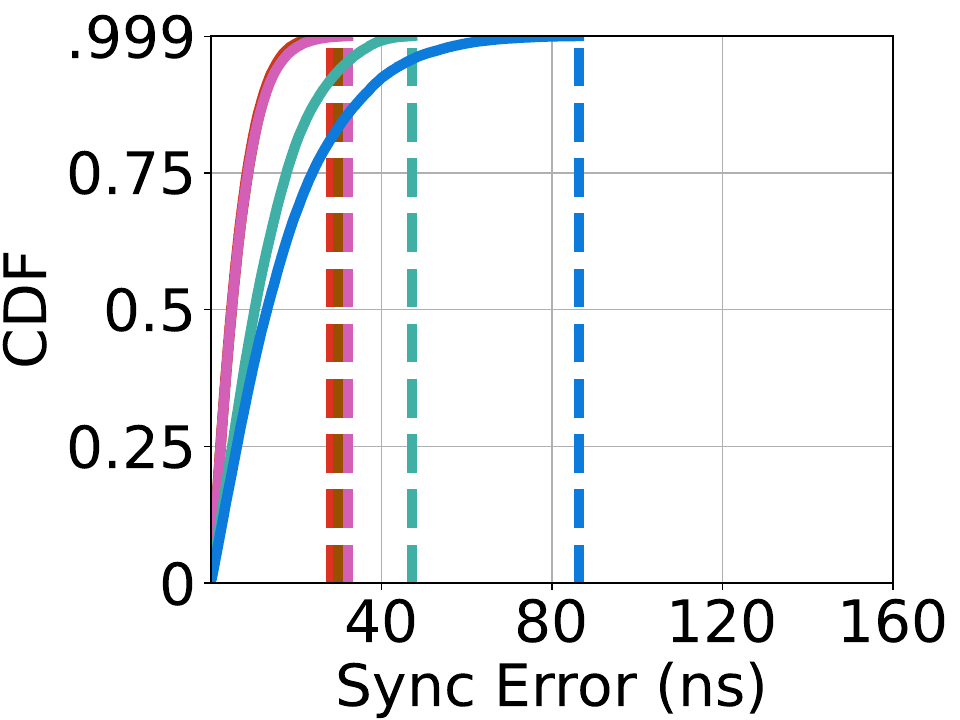}
        \sfigcap{5\% ToR failures}\label{fig:sim_error_node}
    \end{subfigure}
    \begin{subfigure}{0.245\textwidth}
        \includegraphics[width=\linewidth]{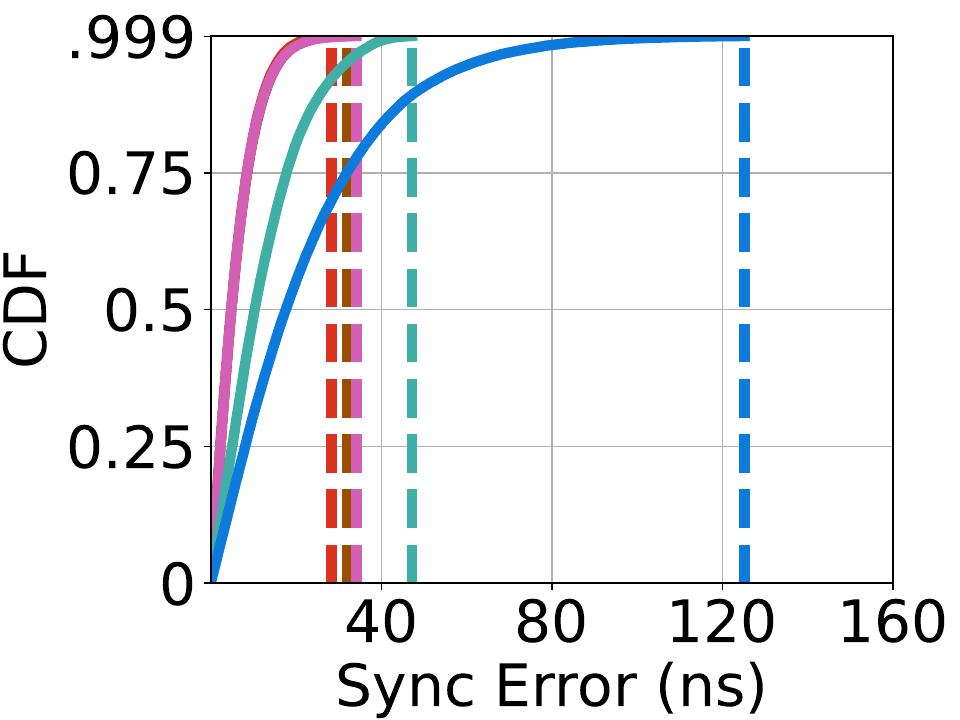}
        \sfigcap{5\% link failures}\label{fig:sim_error_link}
    \end{subfigure}
    \begin{subfigure}{0.245\textwidth}
        \includegraphics[width=\linewidth]{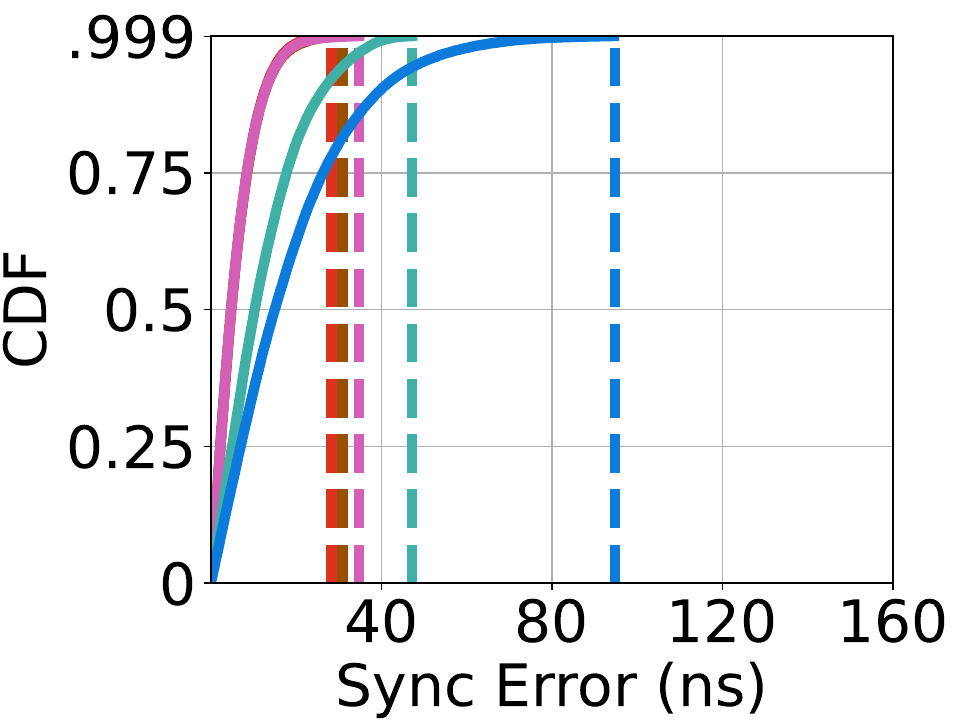}
        \sfigcap{1/12 OCS failure (1 failed link per ToR)}\label{fig:sim_error_ocs}
    \end{subfigure}
    \begin{subfigure}{0.245\textwidth}
        \includegraphics[width=\linewidth]{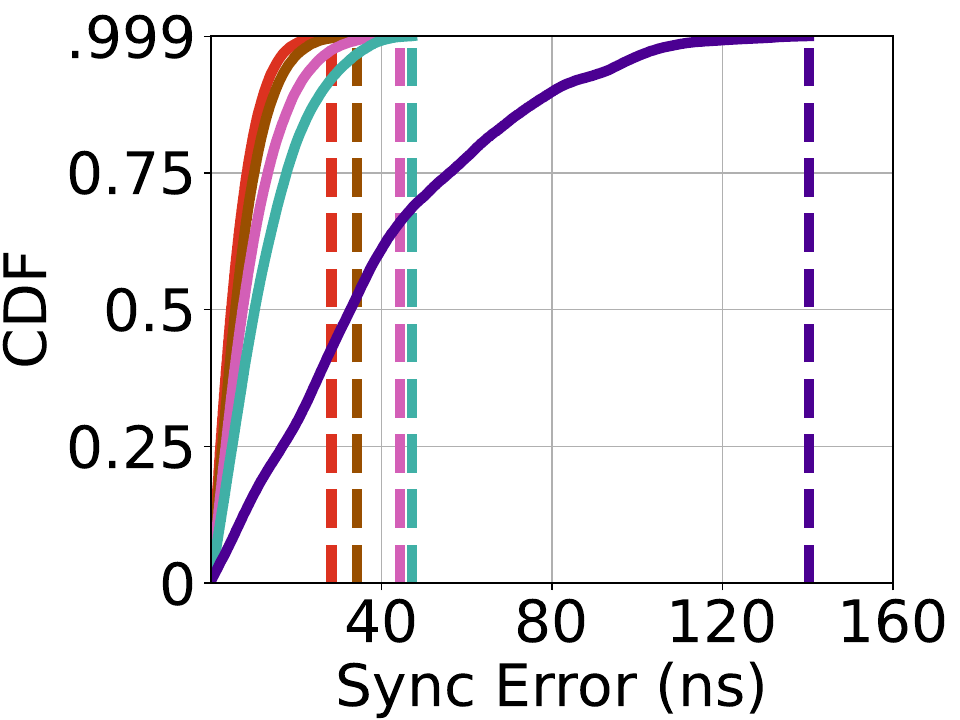}
        \sfigcap{Cooling failure}\label{fig:sim_error_cooling}
    \end{subfigure}
    \figcap{Sync errors of \name  vs. Sundial under four types of failures. 
    }
    \label{fig:sim_error}
\end{figure*}

\para{Q3: How effective are the optimization techniques?}

We have designed single-message sync ($\S$\ref{sec:single-msg}) to support short time slices and timestamp correction ($\S$\ref{sec:hop-error-comp}) to improve sync accuracy. Here, we evaluate their effectiveness.

\paralit{\textbf{Relaxation of slice duration with single-message sync.}}
In Fig.~\ref{fig:sm-sync}, we test the triple-message and single-message sync mechanisms under 10$\mu s$ and 1$\mu s$ time slices, with 40\% background traffic as in Fig.~\ref{fig:bg_traffic}. Triple-message sync requires at least two processing delays and three propagation delays to finish synchronization (Fig.~\ref{fig:ptp}),
i.e., 5.72$\mu$s on our testbed, whereas single-message sync contains no processing delay and only one propagation delay, i.e., 15ns delay in total.

As the figure shows, single-message sync can relax the time slice duration to 1$\mu s$. In Fig.~\ref{fig:sm-sync-10us}, both triple-message and single-message sync can fit into the 10$\mu s$ slice duration, achieving sync accuracy under 11ns. For 1$\mu s$ slices in Fig.~\ref{fig:sm-sync-1us}, single-message sync can still complete within one time slice, but triple-message sync cannot. The unfinished messages will be sent out in the next time slice via a multi-hop path~\cite{RotorNet, Opera}, introducing errors in the delay measurement because of the delay asymmetry. Therefore, single-message sync maintains the same level of accuracy, while the sync errors in triple-message sync increase to 50ns.

\paralit{\textbf{Sync error reduction with timestamp correction.}}
Recall in Fig.~\ref{fig:sync_system_error} we present the sync error distributions among 4 ports, where we observe that timestamp correction reduces the median sync errors, but does not change the range of the sync error variance. This is because we only compensate the intrinsic errors caused by inaccurate timestamping, but cannot eliminate errors incurred by other sources, such as artifacts of the sync protocol. We repeat this experiment but enlarge its coverage to all 32 ports of the switch. 
Fig.~\ref{fig:compensation_bar} presents the absolute median sync errors before and after correction for 12 randomly sampled port pairs. The results are consistent with Fig.~\ref{fig:sync_system_error}: 
the worst sync error before timestamp correction is 53ns, which decreases to 7ns after correction.

\para{Q4: Is \name robust to diverse failures?}

We compare \name with Sundial, the most reliable sync protocol designed for fast failure recovery, under four failure scenarios. We bring down 5\% randomly chosen ToRs and links for ToR and link failures. For \ocs failure, we disable one of the 12 \ocses, which is equivalent to one failed link per ToR (Fig.~\ref{fig:dcn_arch}b).
Since the Sundial static \dcn has no OCS, we disconnect one link per ToR for fair comparison.
For cooling failure, we assume 5$^\circ$C temperature increase causing a drift rate change of $\pm$5ppm. As shown in Fig.~\ref{fig:sim_error}, after handling these failures, the 99.9\textsuperscript{th} percentile tail sync errors of \name increase to 30/31/31/34ns, while for Sundial they increase to 86/125/95/140ns. Overall, \name's accuracy under these failures are 2.9$\times$ to 4.1$\times$ better than Sundial.

For ToR, link, and \ocs failures, the post-failure and post-recovery sync errors of \name are similar, indicating \name's robustness to failures even without recovery. When the backup parent ToR is activated, the sync errors of \name are further reduced, almost reaching the level before the failures. The robustness of \name stems from two factors: (1) it constructs multi-trees, enabling it to easily avoid failures and select a new sync parent; (2) it profiles drifts in advance to choose the sync parent with the minimum expected sync error, while also selecting the second minimum as the backup parent. Sundial also pre-computes backup trees, but for static spanning trees, the backup tree usually has a higher depth than the original one, resulting in higher sync errors.

For the cooling failure case, 
\name's sync error grows to 44ns from 28ns once the cooling malfunctions. After re-profiling and re-calculating the sync plan, the accuracy is back to 34ns. The re-profiling and re-distribution of sync plan are done live,
so the network can operate normally without interruption ($\S$\ref{sec:profiling}). Sundial has no recovery mechanism for cooling failures, and the sync error expands from 47ns to 140ns, similar to the scenario of high drift rate in Fig.~\ref{fig:drift_rate}.

We demonstrate the failure recovery process of \name and Sundial with a link failure case study in Appx.~\ref{sec:appx_recovery_process}.

%% file: conclusion.tex
\vspace{0.20in}
\section{Conclusion}\label{sec:conclusion}
\vspace{0.10in}

We have introduced \name as the first nanosecond-precision synchronization solution for optical \dcns general to various optical architectures. Our key designs such as drift-aware multi-trees, single-message sync, and TX timestamp correction are highly unconventional, which are largely due to the unique characteristics of optical \dcns, but also add interesting design points to the problem space.

As we have observed in $\S$\ref{sec:eval}, most of the performance gains of \name come from the drift-aware multi-trees, which shake the ground of sync protocols in traditional \dcns that the clock propagation tree must be static, aligning with the underlying network topology, and drift-agnostic. And we have shown in the experiments that the benefits of efficient trees can surpass the impact of higher sync frequency. Admittedly, drift and delay profiling in traditional \dcns is more challenging, because of the high interference from the electrical network stack and lack of topology
reconfigurability to enumerate the connections. Nonetheless, Graham has characterized drifts~\cite{graham} and DTP has measured delays reliably in the physical layer~\cite{dtp}. It is not hard to imagine a drift-aware sync scheme for traditional \dcns, not based on exhaustive port-wise profiling like \name, but an approximation from a reasonable understanding of drift and delay properties. Similarly, even without the physically reconfiguring optical network, it is not impossible to build multi-trees in traditional \dcns. Essentially, the backup tree in Sundial has set the first step towards that.

We have defined the in-band synchronization problem for optical \dcns, and we believe \name is by no means the ultimate solution to it. We hope to inspire follow-up work with our released code and spark discussions in the broader context of general \dcns and other network environments.

%% file: appendix.tex
\section*{Appendix}
\appendix

\section{Tree-Building Algorithm}\label{sec:alg}

The clock propagation trees form the sync plan, which specifies which ToR should sync with which other ToR in which time slice. Alg.~\ref{alg:sync_plan} shows the tree-building algorithm for sync plan generation, which takes in the fixed optical schedule of circuit connections and determines sync parents for the ToRs based on their expected sync errors. It runs periodically, each time computing the sync plan for $M$ optical cycles ahead and loading them onto the ToRs (\textit{Lines 3, 10, 11}).

In each time slice of the computation batch, every ToR is connected with a set of other ToRs through the optical circuits. The one with the minimum expected error becomes the potential sync parent for the ToR (\textit{Line 5}).
The ToR is only synchronized if the potential parent's clock is more accurate (\textit{Lines 6--7}). After the sync, the synchronized ToR inherits the expected error from the parent, and its expected error for the next time slice is this inherited error plus the profiled drift for a time slice (\textit{Line 8}). 
Otherwise, if the ToR is not synchronized, its expected error for the next time slice only grows by the drift amount (\textit{Line 9}).

\begin{algorithm}[h]
  \footnotesize
  \caption{Sync Plan Generation}
  \label{alg:sync_plan}
  \begin{algorithmic}[1]
  \Require
  \Statex $M$: frequency of sync plan generation, once every $M$ optical cycles
    \Statex $N$: number of time slices per optical cycle, $U$: time slice duration
    \Statex $S$: the set of $N$ ToRs $\{s_0, s_1, ..., s_{N-1}\}$, $s_0$ is the master ToR
    \Statex $E_{i}^{t}$: expected sync error of $s_i$ relative to $s_0$ before entering time slice $t$
    \Statex ${D}_{i}$: profiled drift error of $s_i$ relative to $s_0$ for one time slice
    \Statex $R_{i}^{t}$: the set of ToRs that $s_i$ connects to in time slice $t$, derived from the cyclic optical schedule
  \Ensure
    \Statex $P$: sync plan, as a set $\{\langle t, s_j \rightarrow s_i\rangle\}$ where $s_i$ syncs with $s_j$ in time slice $t$
    \vspace{2pt}
    \State $E_{0}^{0} \leftarrow 0$; $E_{i}^{0} \leftarrow \infty, 1 \leq i < N$
    \LeftComment{Initialize expected sync errors}

    \Loop
            \For{$t$ in next $M\times N$ time slices} \LeftComment{Compute for a new batch}
            \For{$s_{i}$ in $S$}
                \State{$r=$ $arg$ $min_{j} (E^{t}_{j}), j \in R_{i}^{t}$} \LeftComment{Potential parents}
                \If{$E^{t}_{r} < E^{t}_{i}$ or $r==0$} %
                \State{$P.add(\langle t, s_r \rightarrow s_{i}\rangle)$ \LeftComment{Only sync with better clock}}
                \State{$E_{i}^{t+1} = E_{r}^{t} + {D}_{i}$  \LeftComment{Carry over parent's sync error}}
                \Else{ $E_{i}^{t+1} = E_{i}^{t} + {D}_{i}$ \LeftComment{Not synced, just add drift}}
                \EndIf
            \EndFor
        \EndFor
        \State{Load $P$ to ToRs}
        \State{$sleep(M\times N \times U)$}
    \EndLoop

  \end{algorithmic}
\end{algorithm}

\section{Supplementary of Timestamp Correction}\label{sec:correction_supp}

\vspace{-0.05in}
{\footnotesize
\begin{equation}\tag{7}\label{eqn:c_not_exist}
\begin{split}
delay'_{AC} - delay'_{BC} - (delay_{AC} - delay_{BC}) = \frac{e^{A}-e^{B}}{2} = \Delta_{AB} 
\end{split}
\end{equation}
}

\vspace{-0.10in}
{\footnotesize
\begin{equation}\tag{8}\label{eqn:c_not_exist_2}
\begin{split}
delay'_{AC} - delay'_{BC} - 5\times(l_{AC} - l_{BC})= \frac{e^{A}-e^{B}}{2} = \Delta_{AB}
\end{split}
\end{equation}
}

In the case where $delay_{AC}$ is not equal to $delay_{BC}$, we combine the equations in Eqn.~\ref{eqn:torc} and derive Eqn.~\ref{eqn:c_not_exist}. By substituting the delays with fiber lengths ($l_{AC}$ and $l_{BC}$), we show how to calculate $\Delta_{AB}$ in Eqn.~\ref{eqn:c_not_exist_2}. The unit of the fiber length is meter, and coefficient 5 in Eqn.~\ref{eqn:c_not_exist} means the propagation speed for optical signals is 5ns/m.

\section{Drift measurement}\label{sec:appx_drift}

\begin{figure}[h]

    \begin{subfigure}{0.49\linewidth}
        \includegraphics[width=\linewidth]{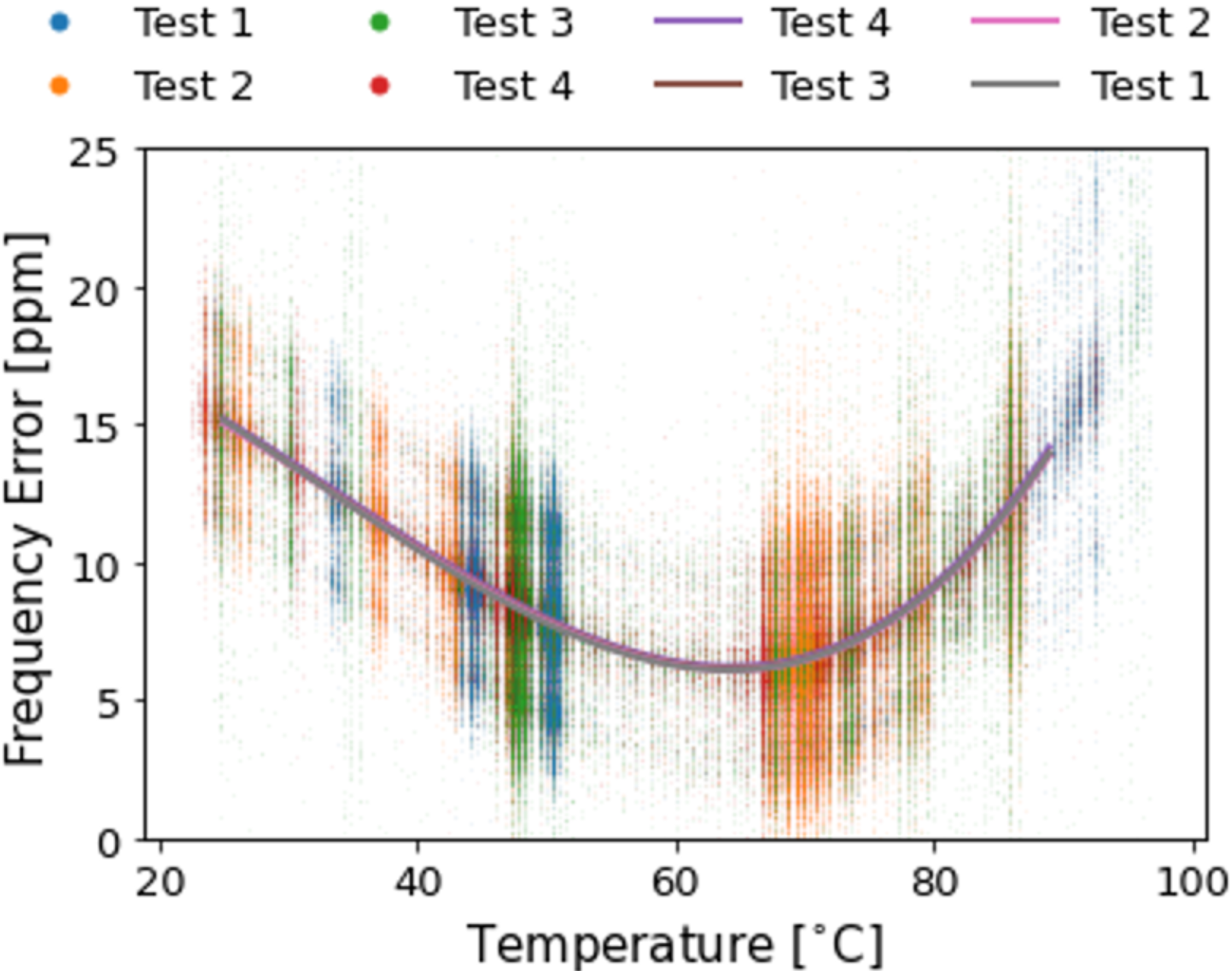}
        \sfigcap{Fig.~8 in the Graham paper~\cite{graham}.}\label{fig:graham_drift}
    \end{subfigure}
    \begin{subfigure}{0.49\linewidth}
        \includegraphics[width=\linewidth]{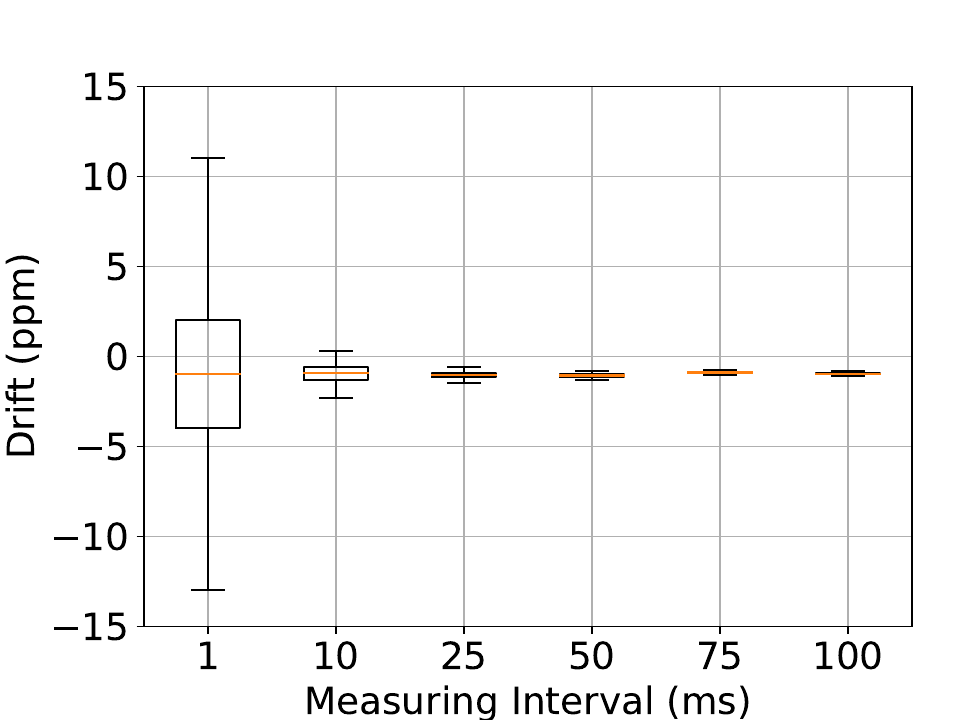}
        \sfigcap{Switch 1.}\label{fig:sw1_drift}
    \end{subfigure}
    \begin{subfigure}{0.49\linewidth}
        \includegraphics[width=\linewidth]{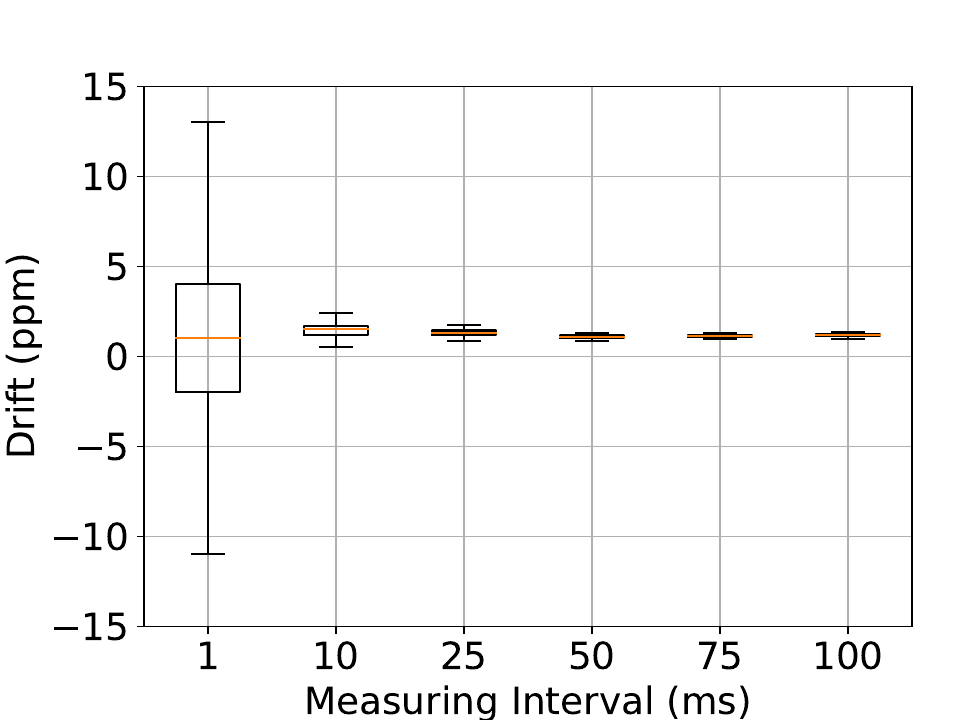}
        \sfigcap{Switch 2.}\label{fig:sw2_drift}
    \end{subfigure}
    \begin{subfigure}{0.49\linewidth}
        \includegraphics[width=\linewidth]{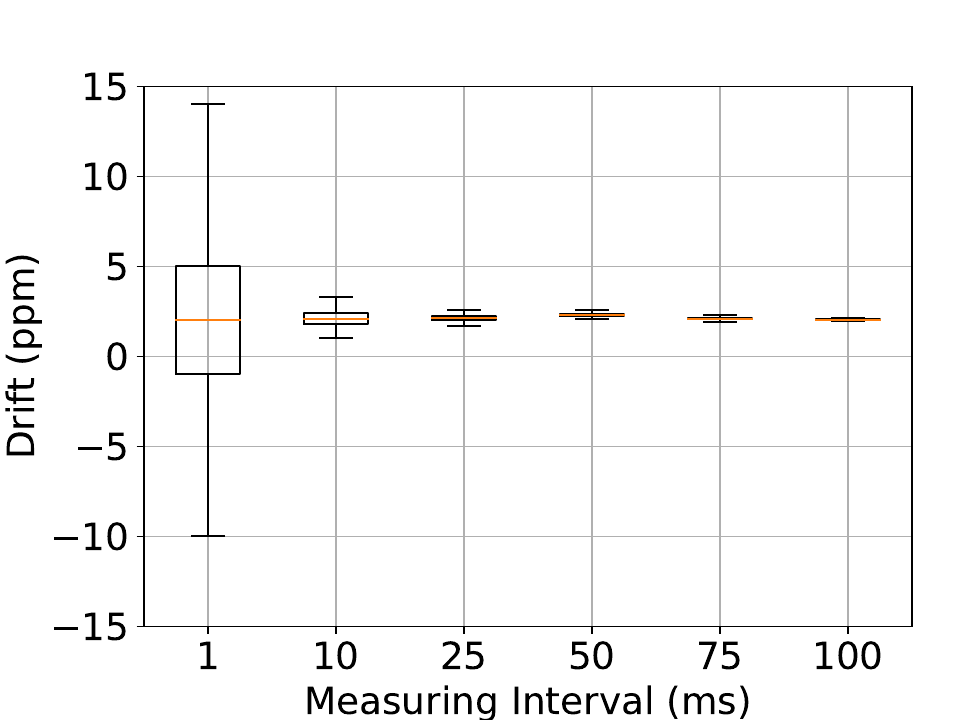}
        \sfigcap{Switch 3.}\label{fig:sw3_drift}
    \end{subfigure}
    \figcap{(a) Graham's drift measurements. (b)-(d) Our drift measurements of three switches against the master switch with different measurement intervals.}\label{fig:drift_measurement}
        
\end{figure}

Fig.~\ref{fig:drift_measurement}a shows Graham's four drift measurements on the same device, indicating the drift expectation is stable across tests.
As presented in Fig.~\ref{fig:drift_measurement}b-d, we validate this finding on three Tofino2 switches in our lab. The drift median per switch is stable across measurements of different measuring intervals, and the variance range for 1ms measuring intervals is consistent with Graham.

\section{Packet Generator Sync}\label{sec:pkg-gen-sync}

The packet generator for generating sync messages needs to be started after the preparatory phase and updated per re-sync to be globally in sync with the optical schedule. Below, we explain how to accomplish this task with the clock offset from \name.

The clock offset is stored in an SRAM-based register on the programmable ToR. We start a \textit{kickoff packet generator} with the minimal packet interval, which per our tests is consistently within 10ns. The absolute time of the first time slice, which is globally agreed upon, is a parameter that can be set by the control plane program. Every packet from the kickoff packet generator reads the offset value and checks if the first time slice has arrived, by adding the offset to the ingress timestamp as the local data plane clock time. On detecting the start of the first time slice, the P4 program signals the control plane to turn off the kickoff packet generator to save packet generation and processing resources on the switch.

Intel Tofino2 allows packet generators to be configured by the control plane and later started in the data plane, triggered by an ingress packet~\cite{lee2020tofino2}. We define a pair of identical \textit{rotation packet generators} for maintaining the correct ticking frequency of the main packet generator in $\S$\ref{sec:enforcement}, whose packet interval is the time slice duration. One such generator is started by the last packet from the kickoff packet generator, which marks the start of the first time slice, with a maximum delay of 10~ns (based on minimal packet interval).
We thus bootstrap the running of the main packet generator.

We adjust the main packet generator only after the clock drift has exceeded a pre-defined threshold.
We store the clock offset used for the current time slice in another register. Every sync message reads this register to check the offset change. When the threshold is reached, this sync message restarts the kickoff packet generator, which now monitors the new clock offset to determine the start of the next time slice, derived from the start time of the first time slice, the time slice duration, and the number of passed time slices.

The packet from the kickoff packet generator that has detected the start of the next time slice, based on the new clock offset, is used to update the main packet generator. It simultaneously disables the current rotation packet generator and enables the other one to follow the updated clock. Our P4 program recalculates the current time slice number with the new clock
to re-start the main packet generator.
Again, the kickoff packet generator is turned off immediately after use. 
This process continues, and the two rotation packet generators are continually swapped, one in use and one idle, to enable live clock adjustments on the data plane.

\section{\name Overhead Analysis}\label{sec:appx_overhead}

\begin{table}[tbp]
\centering
\footnotesize
\tabcap{Sync plan characteristics in a 192-ToR optical \dcn.}
\begin{tabular}{llll}
\hline
\begin{tabular}[c]{@{}l@{}}Slice\\ Time ($\mu$s)\end{tabular} & \begin{tabular}[c]{@{}l@{}}Sync Plan\\ Data (KB)\end{tabular} & \begin{tabular}[c]{@{}l@{}}Max Entries\\ Per ToR\end{tabular} & \begin{tabular}[c]{@{}l@{}}Bandwidth\\ Usage (Mbps)\end{tabular} \\ \hline
1                                                            & 3.1                                                           & 1                                                             & 15.1                                                             \\
50                                                           & 4.7                                                           & 28                                                            & 24.3                                                             \\
100                                                          & 9.9                                                           & 63                                                            & 25.8                                                             \\
300                                                          & 28.4                                                          & 155                                                           & 24.6                                                             \\
500                                                          & 33.1                                                          & 175                                                           & 17.3                                                             \\
1000                                                         & 33.4                                                          & 177                                                           & 8.7                                                              \\ \hline
\end{tabular}\label{tb:overehad}
\end{table}

We measure the overhead of \name in a large-scale optical \dcn setup with 192 ToRs.
The sync planner distributes the sync plans to other ToRs at the preparatory phase and the operational phase.
Table~\ref{tb:overehad} shows that the total data size of the sync plans in the first 50 optical cycles ranges from 3.1KB to 33.4KB. The plan updates for further optical cycles are negligible since we only send the deltas.
Overall, the sync planner's transmission overhead is minimal.

Table~\ref{tb:overehad} also shows that the sync plan takes minimal memory at ToRs.
ToRs need to store at most 177 entries of sync plans and backup plans when the slice duration is 1000$\mu$s, taking 2.8KB of switch memory, which is negligible compared to the memory resources at modern switches~\cite{miao2017silkroad}.

Sync operations take minimal network resources at ToRs.
In our implementation, each sync message is 100B, adding only 8ns buffering for other packets at 100Gbps links.
Sending sync messages takes at most 25.8Mbps bandwidth per ToR which is minimal for modern tens of Tbps switches\cite{tofino2}.
If we leverage the physical layer messages in synchronization, as in \cite{dtp}, the bandwidth usage can be further reduced.

\section{Failure Recovery Case Study}\label{sec:appx_recovery_process}
\begin{figure}[tbp]
    \centering 
    \includegraphics[width=0.50\linewidth]{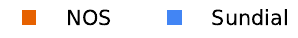}
    \vfil
    \begin{subfigure}[tbp]{0.90\columnwidth}
        \includegraphics[width=\linewidth]{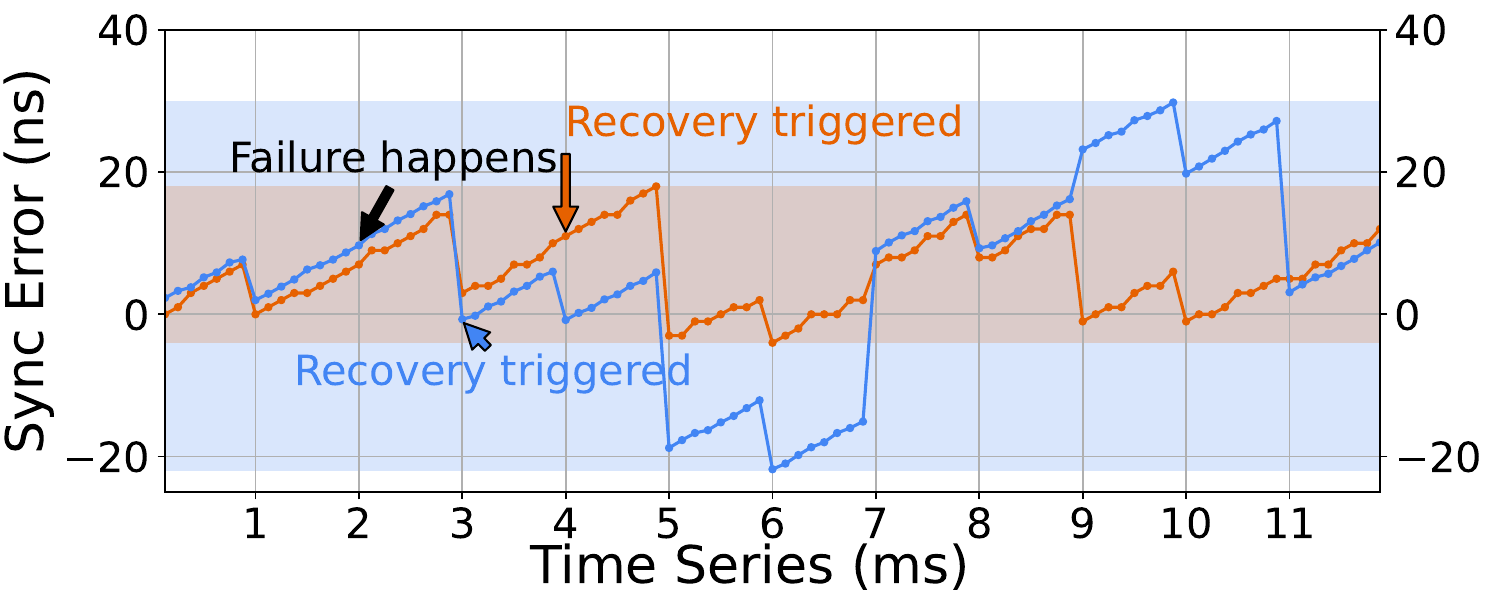}
        \vspace{-1.8em}
        \sfigcap{
        }\label{fig:recovery_toy_example}
    \end{subfigure}
    \hfill
    \begin{subfigure}[tbp]{0.90\columnwidth}
        \includegraphics[width=\linewidth]{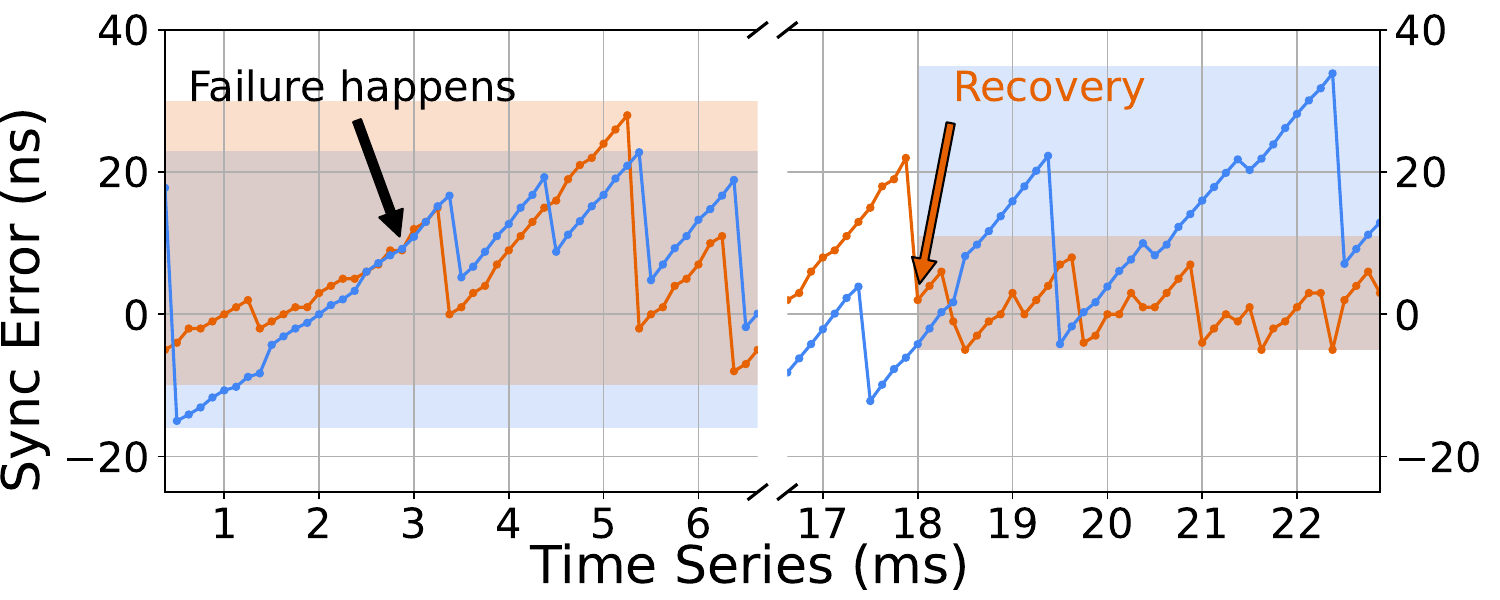}
        \vspace{-1.8em}
        \sfigcap{
        }\label{fig:recovery_toy_example_cooling}
    \end{subfigure}
    \figcap{Recovery of (a) a link failure, (b) a cooling failure.}\label{fig:recovery}
\end{figure}

Fig.~\ref{fig:recovery_toy_example} shows how \name and Sundial conduct recovery for a ToR affected by a link failure. We can see that the recovery in Sundial is triggered earlier than \name. This is because Sundial syncs with only one ToR and so it can find the failure sooner. However, after recovering, Sundial cannot maintain the same sync accuracy before failure. Sundial's sync accuracy increases from 18ns to 30ns. In contrast, even though the recovery in \name is triggered later, it maintains nearly the same sync accuracy before and after the failure.

Fig.~\ref{fig:recovery_toy_example_cooling} depicts the recovery processes of \name and Sundial after a cooling failure happens. Once the cooling failure occurs, the drift errors become higher, leading to the increment of the sync errors.
\name conducts re-profiling and applies a new sync plan, reducing the sync accuracy further to 11ns due to higher sync frequency in the new sync plan to react to the higher drift. 
Sundial does not handle cooling failures, and the sync error increases from 23ns to 35ns.